\documentclass{aa}
\usepackage{graphicx}
\def\deg{$^\circ$\hspace{-1.7mm}.\hspace{0.3mm}}  
\def\as{\arcsec\hspace{-1.2mm}.\hspace{0.3mm}} 
\def\am{\arcmin\hspace{-1.2mm}.\hspace{0.3mm}}

\begin{document}
   \title{$UBVRI$ Night sky brightness during sunspot maximum at ESO-Paranal
   \thanks{Based on observations collected at ESO-Paranal.}}

   \author{F. Patat}
   \offprints{F. Patat;  \email{fpatat@eso.org}}

   \institute{European Southern Observatory, K.Schwarzschild Str.2,
              85748 - Garching - Germany}

   \date{Received October 24, 2002; accepted December 19, 2002}

\abstract{
In this paper we present and discuss for the first time a large data set 
of $UBVRI$ night sky brightness measurements collected at ESO-Paranal from 
April 2000 to September 2001. A total of about 3900 images obtained on 174
different nights with FORS1 were analysed using an automatic algorithm 
specifically designed for this purpose. This led to the construction of
an unprecedented database that allowed us to study in detail a number of 
effects such as differential zodiacal light contamination, airmass dependency,
daily solar activity and moonlight contribution. Particular care was 
devoted to the investigation of short time scale variations and micro-auroral
events. The typical dark time night sky brightness values found for Paranal 
are similar to those reported for other astronomical dark sites at a similar 
solar cycle phase. The zenith-corrected values averaged over the whole period
are 22.3, 22.6, 21.6 20.9 and 19.7 mag arcsec$^{-2}$ in $U, B, V, R$ and $I$ 
respectively. In particular, there is no evidence of light pollution either
in the broadband photometry or in the high-airmass spectra we have analysed.
Finally, possible applications for the exposure time calculators are 
discussed.
  
\keywords{atmospheric effects -- site testing -- light pollution --
techniques: photometric}
}

\titlerunning{Night sky brightness at ESO-Paranal}
\authorrunning{Ferdinando Patat}

\maketitle

%
%________________________________________________________________

\section{\label{sec:intro} Introduction}

The night sky brightness, together with number of clear nights, seeing, 
transparency, photometric stability and humidity, are some of the most
important parameters that qualify a site for front-line 
ground--based astronomy. While there is almost no way to control
the other characteristics of an astronomical site, the sky brightness
can be kept at its {\it natural} level by preventing 
light pollution in the observatory areas. This can be achieved
by means of extensive monitoring programmes aimed at detecting any possible
effects of human activity on the measured sky brightness.
 
For this purpose, we have started an automatic survey of the $UBVRI$ night sky
brightness at Paranal with the aim of both getting for the first time values 
for this site and building a large database. The latter is a fundamental 
step for the long term trend which, given the possible growth of human 
activities around the observatory, will allow us to check 
the health of Paranal's sky in the years to come.
  
The ESO-Paranal Observatory is located on the top of Cerro Paranal in the 
Atacama Desert in the northern part of Chile, one of the driest areas on 
Earth. Cerro Paranal (2635 m, 24$^\circ$ 40$^\prime$ S, 
70$^\circ$ 25$^\prime$ W) is at about 108 km S of Antofagasta 
(225,000 inhabitants; azimuth 0\deg2), 280 km SW from Calama 
(121,000 inhabitants; azimuth 32\deg3), 152 km WSW from La Escondida
(azimuth 32\deg9), 23 km NNW from a small mining
plant (Yumbes, azimuth 157\deg7) and 12 km inland from the Pacific Coast.  
This ensures that the astronomical observations to be carried out there 
are not disturbed by adverse human activities like dust and light from 
cities and roads. Nevertheless, a systematic monitoring of the
sky conditions is mandatory in order to preserve the high site quality
and to take appropriate action, if the conditions are proven to
deteriorate. Besides this, it will also set the stage for the study of 
natural sky brightness oscillations, both on short and long time scales, 
such as micro-auroral activity, seasonal and sunspot cycle effects.

The night sky radiation has been studied by several authors, starting with
the pioneering work by Lord Rayleigh in the 1920s. 
For thorough reviews on this subject the reader is referred to the classical 
textbook by Roach \& Gordon (\cite{roachgordon}) and the recent extensive 
work by Leinert et al. (\cite{leinert98}), which explore a large number
of aspects connected with the study of the night sky emission.
In the following, we will give a short introduction to the subject,
concentrating on the optical wavelengths only.

\begin{figure}
\centering
\resizebox{\hsize}{!}{\includegraphics{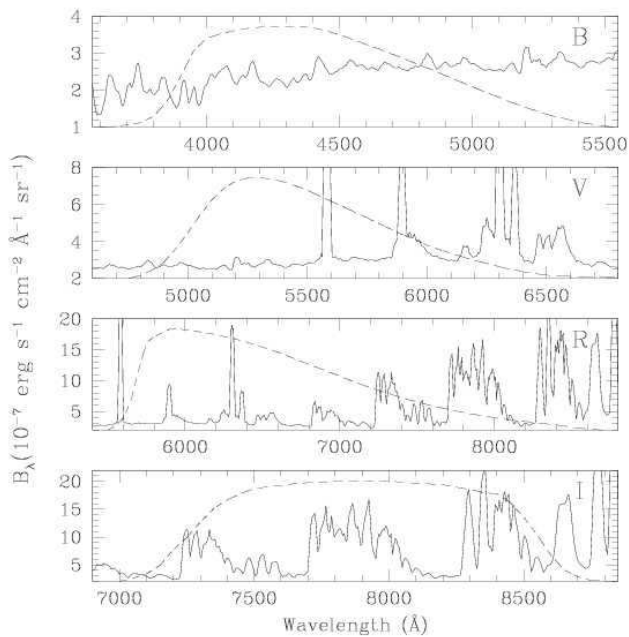}}
\caption{\label{fig:skyspectra} Night sky spectrum obtained at Paranal
on February 25, 2001 02:38UT in the spectral region covered by 
$B$, $V$, $R$ and $I$ passbands (from top to bottom). The original 
FORS1 1800 seconds frame was taken at 1.42 airmasses with a long slit of 
1$^{\prime\prime}$ and grism 150I, which provide a  resolution of about 
22 \AA\/ (FWHM). The dashed lines indicate the passband response curves.
Flux calibration was achieved using the spectrophotometric standard star 
Feige~56 (Hamuy et al. 1992) observed during the same night. The absence
of an order sorting filter probably causes some second order overlap
at wavelengths redder than 6600 \AA.}
\end{figure}

The night sky light as seen from ground is generated by several sources, 
some of which are of extra-terrestrial nature (e.g. unresolved 
stars/galaxies, diffuse galactic background, zodiacal light) and 
others are due to atmospheric phenomena (airglow and auroral activity
in the upper Earth's atmosphere). In addition to these {\it natural} 
components, human activity has added an extra source, namely the 
artificial light scattered by the troposphere, mostly in the form of
Hg-Na emission lines in the blue-visible part of the 
optical spectrum (vapour lamps) and a weak continuum (incandescent lamps).
While the extra-terrestrial components vary only with the position on
the sky and are therefore predictable, the terrestrial ones are known 
to depend on a large number of parameters (season, geographical position, 
solar cycle and so on) which interact in a largely unpredictable way.
In fact, airglow contributes with a significant
fraction to the optical global night sky emission and hence its variations 
have a strong effect on the overall brightness.

To illustrate the various processes which contribute to the
airglow at different wavelengths, in Figure~\ref{fig:skyspectra} we have plotted
a high signal-to-noise, flux calibrated night sky spectrum obtained at Paranal 
on a moonless night (2001, Feb 25) at a zenith distance of 45$^\circ$ ,
about two hours after the end of evening astronomical twilight. In the 
$B$ band the spectrum is rather featureless and it is characterised by the 
so called airglow pseudo-continuum, which arises in layers at a height of 
about 90-100 km (mesopause). This actually extends from 4000\AA\/ to 
7000\AA\/ and its intensity is of the order of 3$\times$10$^{-7}$ erg 
s$^{-1}$ cm$^{-2}$ \AA$^{-1}$ sr$^{-1}$ at 4500\AA. All visible emission 
features, which become particularly marked below 4000\AA\/
and largely dominate the $U$ passband (not included in the plots),
are due to Herzberg and Chamberlain O$_2$ bands (Broadfoot \& Kendall 1968).
In light polluted sites, this spectral region is characterised by the
presence of Hg I (3650, 3663, 4047, 4078, 4358 and 5461\AA) and NaI (4978, 
4983, 5149 and 5153\AA) lines (see for example Osterbrock \& Martel, 1992) 
which are, if any, very weak in the spectrum of Figure~\ref{fig:skyspectra} 
(see Sec.~\ref{sec:discuss} for a discussion on light pollution at 
Paranal). Some of these lines are clearly visible in spectra taken, for 
example,  at La Palma (Benn \& Ellison 1998, Figure~1) and Calar Alto 
(Leinert et al. 1995, Figs.~7 and 8).

The $V$ passband is chiefly dominated by [OI]5577\AA\/ and to a lesser
extent by NaI~D and [OI]6300,6364\AA\/ doublet. In the spectrum of 
Figure~\ref{fig:skyspectra} the relative contribution to the total flux of these
three lines is 0.17, 0.03 and 0.02, respectively. Besides 
the aforementioned pseudo-continuum, several OH Meinel vibration-rotation 
bands are also present in this spectral window (Meinel 1950); in particular,
OH(8-2) is clearly visible on the red wing of NaI~D lines and  OH(5-0), 
OH(9-3) on the blue wing of [OI]6300\AA. All these features are known to be 
strongly variable and show independent behaviour (see for example the 
discussion in Benn \& Ellison 1998), probably due to the fact that 
they are generated
in different atmospheric layers (Leinert et al. 1998 and references 
therein). In fact, [OI]5577\AA, which is generally  the brightest emission 
line in the optical sky spectrum, arises in layers at an altitude of 90 km, 
while [OI]6300,6364\AA\/ is produced at 250-300 km. 
The OH bands are emitted by a layer at about 85 km, while the Na ID is 
generated at about 92 km, in the so called Sodium-layer which is used by 
laser guide star adaptive optic systems.
In particular, [OI]6300,6364\AA\/ shows a marked and complex dependency on 
geomagnetic latitude  which turns into different typical line intensities at 
different observatories (Roach \& Gordon \cite{roachgordon}). Moreover, this doublet undergoes
abrupt intensity changes (Barbier 1957); an example of such an event is 
reported and discussed in Sec.~\ref{sec:discuss}.

In the $R$ passband, besides the contribution of NaI~D and [OI]6300,6364\AA,
which account for 0.03 and 0.10 of the total flux in the spectrum of
Figure~\ref{fig:skyspectra}, strong OH Meinel bands like OH(7-2), OH(8-3), 
OH(4-0), OH(9-4) and OH(5-1) begin to appear, while the pseudo-continuum 
remains constant at about 3$\times$10$^{-7}$ erg s$^{-1}$ 
cm$^{-2}$ \AA$^{-1}$ sr$^{-1}$. Finally, the $I$ passband is dominated by the
Meinel bands OH(8-3), OH(4-0), OH(9-4), OH(5-1) and OH(6-2); the broad
feature visible at 8600-8700\AA, and marginally contributing to the
$I$ flux, is the blend of the R and P branches of O$_2$(0-1) (Broadfoot 
\& Kendall 1968).

Several sky brightness surveys have been performed at a number of
observatories in the world, most of the time in $B$ and $V$ passbands
using small telescopes coupled to photo-multipliers.
A comprehensive list of published data is given by Benn \& Ellison 
(\cite{benn}). 
All authors agree on the fact that the dark time sky brightness shows 
strong variations within the same night on the time scales 
of tens of minutes to hours. This variation is 
commonly attributed to airglow fluctuations. Moreover, as first pointed 
out by Rayleigh (\cite{rayleigh}), the intensity of the [OI]5577\AA\/
line depends on the solar activity. Similar results were found for other
emission lines (NaI~D and OH) by Rosenberg \& Zimmerman (\cite{rosenberg}).
Walker (\cite{walker88b}) found that $B$ and $V$ sky brightness is well
correlated with the 10.7 cm solar radio flux and reported a range of 
$\sim$ 0.5 mag in $B$ and $V$ during a full sunspot cycle.
Similar values were found by Krisciunas (\cite{krisc90}), Leinert et 
al. (\cite{leinert95}) and Mattila et al. (\cite{attila}), so that the effect
of solar activity is commonly accepted (Leinert et al. 1998).
For this reason, when comparing sky brightness measurements, one should
also keep in mind the time when they were obtained with respect to the
solar cycle, since the difference can be substantial.
A matter of long debate has been the so-called {\it Walker effect},
named after Walker (\cite{walker88b}), who reported a steady exponential
decrease of  $\sim$0.4 mag in the night sky brightness during the first six 
hours following the end of twilight. This finding has been questioned by 
several authors. We address this issue in detail later 
(Sec.~\ref{sec:variations} and Appendix \ref{sec:walker}).

Here we present for the first time $UBVRI$ sky brightness 
measurements for Paranal, obtained on 174 nights from 2000 April 20 to
2001 September 23 which, to our knowledge, makes it the largest 
homogeneous data set available. Being produced by an automatic procedure,
this data base is continuously growing and it will provide an unprecedented
chance to investigate both the long term evolution of the night sky 
quality and to study in detail the short time scale fluctuations which are
still under debate.
 
The paper is organized as follows. After giving some information on the
basic data reduction procedure in Sec.~\ref{sec:red}, in Sec.~\ref{sec:calib}
we discuss the photometric calibration and error estimates, while
the general properties of our night sky brightness survey are described in 
Sec.~\ref{sec:survey}. 
The results obtained during dark time are then presented in 
Sec.~\ref{sec:dark} and the short time-scale variations are analysed 
in Sec.~\ref{sec:variations}. In Sec.~\ref{sec:moonmod} we compare our data
obtained in bright time with the model by Krisciunas \& Schaefer 
(\cite{krischae}) for the effects of moonlight, while the dependency on 
solar activity is investigated in Sec.~\ref{sec:solar}. In 
Sec.~\ref{sec:discuss} we discuss the results and summarize our conclusions.
Finally, detailed discussions about some of the topics are given in Appendices
A--D.

\section{\label{sec:red} Observations and basic data reduction}

The data set discussed in this work has been obtained with the FOcal 
Reducer/low dispersion Spectrograph (hereafter FORS1), mounted
at the Cassegrain focus of ESO--Antu/Melipal 8.2m telescopes
(Szeifert 2002). The instrument is 
equipped with a 2048$\times$2048 pixels (px) TK2048EB4-1 backside thinned 
CCD and has two remotely exchangeable collimators, which give a projected 
scale of 0\as2 and 0\as1 per pixel (24$\mu$m $\times$ 24$\mu$m). According 
to the used collimator, the sky area covered by the detector is 
6\am8$\times$6\am8 and 3\am4$\times$3\am4, respectively. Most of the 
observations discussed in this paper were performed with the lower resolution
collimator, since the higher resolution is used only to exploit excellent 
seeing conditions (FWHM$\leq$0\as4).

In the current operational scheme, FORS1 is offered roughly in equal fractions
between visitor mode (VM) and
service mode (SM). While VM data are immediately released to the visiting
astronomers, the SM data are processed by the FORS-Pipeline and then 
undergo a series of quality control (QC) checks before being delivered 
to the users. In particular, the imaging frames are bias and flat-field 
corrected and the resulting products are analysed in order
to assess the accuracy of the flat-fielding, the image quality and so on. 
The sky background measurement was experimentally introduced in the QC 
procedures starting with April 2000. Since then, each single
imaging frame obtained during SM runs is used to measure the sky 
brightness. During the
first eighteen months of sky brightness monitoring, more than 4500 frames
taken with broad and narrow band filters have been analysed.

As already mentioned, all imaging frames are automatically bias and flat 
field corrected by the FORS pipeline.
This is a fundamental step, since in the case of imaging, the FORS1 detector
is readout using four amplifiers which have different gains. The bias and 
flat field correction remove the four-port structure to within $\sim$1
electron. This has to be compared with the RMS read-out noise (RON), which is
5.5 and 6.3 electrons in the high gain and low gain modes respectively.
Moreover, due to the large collecting area of the telescope, FORS1 imaging
frames become sky background dominated already after less than two minutes. 
The only significant exception is the $U$ passband, where background
domination occurs after more than 10 minutes (see also Table~\ref{tab:rates}).
The dark current 
of FORS1 detector is $\sim$2.2 $\times$ 10$^{-3}$ e$^{-}$ s$^{-1}$ px$^{-1}$ 
(Szeifert \cite{szeifert}) and hence its contribution to the background can 
be safely neglected.   

\begin{table}
\centering
\caption{\label{tab:rates} Typical background count rates 
expected in FORS1 (SR) images during dark time and at zenith.
The last column reports the time required to have a background photon shot
noise three times larger than the maximum RON (6.3 $e^-$),
which corresponds to a 5\% contribution of RON to the global noise.}
\begin{tabular}{ccc}
\hline \hline
Passband & Count Rate                   & $t_{3}$\\
         & (e$^{-}$ px$^{-1}$ s$^{-1}$) & (s)\\
\hline
 U & 0.5  & 714 \\
 B & 3.8  & 94 \\
 V & 15.8 & 23 \\
 R & 26.7 & 13 \\
 I & 32.1 & 11 \\
\hline
\end{tabular}
\end{table}

Since the flat fielding is performed using twilight sky flats,
some large scale gradients are randomly introduced by the flat
fielding process; maximum peak-to-peak residual deviations are of the order 
of 6$\%$.  Finally, small scale features are very well removed, the only 
exceptions being some non-linear pixels spread across the detector.

The next step in the process is the estimate of the sky background. 
Since the science frames produced by FORS1 are, of course, not necessarily 
taken in {\it empty} fields, the background measurement requires 
a careful treatment. For this purpose we have designed a specific algorithm,
which is presented and discussed in Patat (\cite{patat}). The reader is
referred to that paper for a detailed description of the problem and the
technique we have adopted to solve it.

\section{\label{sec:calib} Photometric calibration and global errors}

Once the sky background $I_{sky}$ has been estimated, the flux per
square arcsecond and per unit time is given by
$f_{sky} = I_{sky}/(t_{exp} \; p^2)$,
where $p$ is the detector's scale (arcsec pix$^{-1}$) and $t_{exp}$ is 
the exposure time (in seconds). The instrumental sky surface brightness 
is then defined as

\begin{equation}
\label{eq:skyb}
m_{sky} = -1.086\;ln(I_{sky}) + 2.5\;log(p^2 \; t_{exp})
\end{equation}

with $m_{sky}$ expressed in mag arcsec$^{-2}$.
Neglecting the errors on $p$ and $t_{exp}$, one can compute the error 
on the sky surface brightness as 
$\delta_{b_{sky}} \simeq \delta_{I_{sky}}/I_{sky}$.
This means, for instance, that an error of 1\% on $I_{sky}$ produces an 
uncertainty of 0.01 mag arcsec$^{-2}$ on the final instrumental surface 
brightness estimate. While in previous photoelectric sky brightness surveys 
the uncertainty on the diaphragm size contributes to the global error in a 
relevant way (see for example Walker 1988b), in our case the pixel scale is 
known with an accuracy of better than 0.03 \% (Szeifert 2002) and the 
corresponding photometric error can therefore be safely neglected.

\begin{figure}
\resizebox{\hsize}{!}{\includegraphics{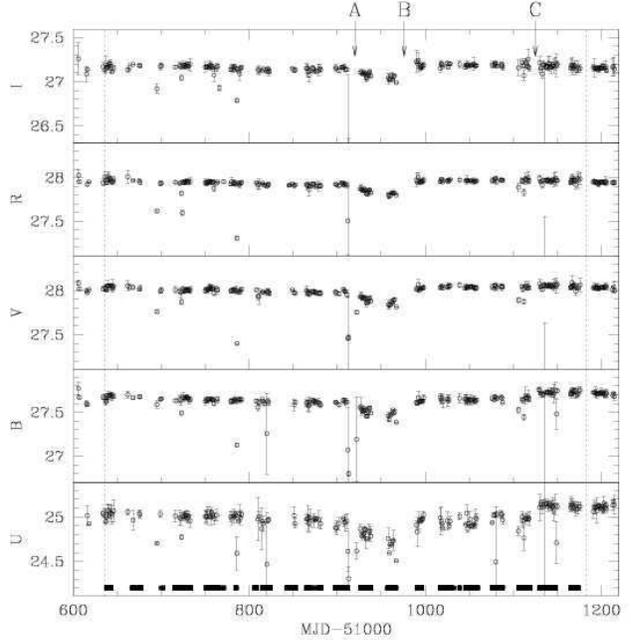}}
\caption{\label{fig:zp} UBVRI photometric zeropoints for FORS1 during
the time range covered by sky brightness measurements presented in this
work (vertical dotted lines). The thick segments plotted
on the lower diagram indicate the presence of sky brightness data, while
the arrows in the upper part of the figure correspond to some relevant
events. A: water condensation on main mirror of UT1-Antu.
B: UT1-Antu main mirror re-aluminisation after the water condensation event. 
C: FORS1 moved from UT1-Antu to UT3-Melipal.  
Plotted zeropoints have been corrected for extinction and colour terms
using average values (see Appendix \ref{sec:photcoeff}).
}
\end{figure}

The next step one needs to perform to get the final sky surface brightness 
is to convert the instrumental magnitudes to the standard $UBVRI$ photometric 
system. Following the prescriptions by Pilachowski et al. (\cite{pila}), 
the sky brightness is calibrated without correcting the measured flux by 
atmospheric extinction, since the effect is actually taking place mostly
{\it in} the atmosphere itself. This is of course not true for the 
contribution coming from faint stars, galaxies and the zodiacal light, which 
however account for a minor fraction of the whole effect, airglow being the
prominent source of night sky emission in dark astronomical sites. 
The reader is referred to Krisciunas (\cite{krisc90}) for a more detailed 
discussion of this point; here we add only that this practically
corresponds to set to zero the airmass of the observed sky area in the
calibration equation. Therefore, if $M_{sky}$, $m_{sky}$ are the calibrated
and instrumental sky magnitudes, $M_*$, $m_*$ are the corresponding values
for a photometric standard star observed at airmass $z_*$, and $\kappa$ is
the extinction coefficient, we have that
$M_{sky}=(m_{sky}-m_*) + \kappa\;z_* + M_* + \gamma \; (C_{sky}-C_*)$. This 
relation can be rewritten in a more general way as 
$M_{sky}=m_0+m_{sky}+\gamma \; C_{sky}$, where $m_0$ is the photometric 
zeropoint in a given passband and $\gamma$ is the colour term in that 
passband for the color $C_{sky}$. For example, in the case of B filter,
this relation can be written as 
$B_{sky}=B_0+b_{sky}+\gamma^B_{B-V} \;(B-V)_{sky}$.

In the case of FORS1, observations of 
photometric standard fields (Landolt 1992) are regularly obtained as part 
of the calibration plan; typically one to three fields are observed during 
each service mode night. The photometric zeropoints were derived from these
observations by means of a semi-automatic procedure, assuming constant 
extinction coefficients and colour terms. For a more detailed discussion on 
these parameters, the reader is referred to Appendix \ref{sec:photcoeff}, 
where we show that this is a reasonable assumption.
Figure~\ref{fig:zp} shows that, with the exception of a few cases,
Paranal is photometrically stable, being the RMS zeropoint fluctuation
$\sigma_{m_0}$=0.03 mag in $U$ and $\sigma_{m_0}$=0.02 mag in all other 
passbands.
Three clear jumps are visible in Figure~\ref{fig:zp}, all basically 
due to physical changes in the main mirror of the telescope. Besides
these sudden variations, we have detected a slow decrease in the efficiency
which is clearly visible in the first 10 months and is most probably
due to aluminium oxidation and dust deposition. The efficiency loss
appears to be linear in time, with a rate steadily decreasing from blue to 
red passbands, being 0.13 mag yr$^{-1}$ in $U$ and  0.05 mag yr$^{-1}$ in $I$.
To allow for a proper compensation of these effects, we have divided the whole
time range in four different periods, in which we have used a linear least 
squares fit to the zero points obtained in each band during photometric nights
only. This gives a handy description of the overall system efficiency which 
is easy to implement in an automatic calibration procedure.

\begin{table*}
\centering
\caption{\label{tab:colors} Typical night sky broad band colours measured
at various observatories.}
\begin{tabular}{llccccl}
\hline \hline
Site       & Year   & $U-B$   & $B-V$& $V-R$ & $V-I$& Reference \\
\hline
Cerro Tololo&1987-8 & $-$0.7  & +0.9 & +0.9  & +1.9 & Walker (1987, 1988a)\\
La Silla   & 1978   & -       & +1.1 & +0.9  & +2.3 & Mattila et al. (1996)\\
Calar Alto & 1990   & $-$0.5  & +1.1 & +0.9  & +2.8 & Leinert et al. (1995)\\
La Palma   & 1994-6 & $-$0.7  & +0.8 & +0.9  & +1.9 & Benn \& Ellison (1998)\\
Paranal    & 2000-1 & $-$0.4  & +1.0 & +0.8  & +1.9 & this work \\
\hline
\end{tabular}
\end{table*}

To derive the colour correction included in the calibration equation one needs
to know the sky colours $C_{sky}$. In principle $C_{sky}$ can be computed
from the instrumental magnitudes, provided that the data which correspond
to the two passbands used for the given colour are taken closely in time.
In fact, the sky brightness is known to have quite a strong time evolution
even in moonless nights and far from twilight (Walker 1988b, Pilachowski et
al. 1989, Krisciunas 1990, Leinert et al. 1998) and using magnitudes obtained
in different conditions would lead to wrong colours. On the other hand, very
often FORS1 images are taken in rather long sequences, which make use of
the same filter; for this very reason it is quite rare to have
close-in-time multi-band observations. Due to this fact and to allow for a 
general and uniform approach, we have decided to use constant sky colours for the 
colour correction. In fact, the colour terms are small, and even large errors on 
the colours produce small variations in the corresponding colour correction.

For this purpose we have used color-uncorrected sky brightness values obtained
in dark time, at airmass $z\leq1.3$ and at a time distance from the twilights
$\Delta t_{twi}\geq2.5$ hours and estimated the typical value as the average.
The corresponding colours are shown in Table~\ref{tab:colors}, where they are
compared with those obtained at other observatories. As one can see, $B-V$
and $V-R$ show a small scatter between different observatories, while $U-B$ 
and $V-I$ are rather dispersed.
In particular, $V-I$ spans almost a magnitude, the value reported for Calar
Alto being the reddest. This is due to the fact that the Calar Alto sky in $I$
appears to be definitely brighter than in all other listed sites.

Now, given the color terms reported in Table~\ref{tab:cterm}, the color 
corrections $\gamma\times C_{sky}$ turn out to be  $-$0.02$\pm$0.02, 
$-$0.09$\pm$0.02, +0.04$\pm0.01$, +0.02$\pm$0.01 and $-$0.08$\pm$0.02 in $U$, 
$B$, $V$, $R$ and $I$ respectively. The uncertainties were estimated from the
dispersion on the computed average colours, which is $\sigma_C\simeq$0.3 for all 
passbands.
We emphasize that this rather large value is not due to measurement errors, 
but rather to the strong intrinsic variations shown by the sky brightness, 
which we will discuss in detail later on. We also warn the reader that the 
colour corrections computed assuming dark sky colours are not necessarily 
correct under other conditions, when the night sky emission is strongly 
influenced by other sources, like Sun and moon. At any rate, colour 
variations of 1 mag would produce a change in the calibrated magnitude of 
$\sim$0.1 mag in the worst case.

Due to the increased depth of the emitting layers, the sky becomes inherently
brighter for growing zenith distances (see for example Garstang 1989, 
Leinert et al. 1998). In order to compare and/or combine together sky brightness 
estimates obtained at different airmasses, one needs to take into account this
effect. The law we have adopted for the airmass compensation and its ability
to reproduce the observed data are discussed in Appendix~\ref{sec:airmdep} 
(see Eq.~\ref{eq:airmdep}).
After including this correction in the calibration equation and neglecting 
the error on $X$, we have computed the global RMS error on the estimated sky 
brightness in the generic passband as follows:

\begin{equation}
\sigma_{M_{sky}}\simeq\sqrt{\delta^2_{m_{sky}}+\sigma^2_{m_0}+\gamma^2\;\sigma^2_C
+C^2\;\sigma^2_\gamma +(X-1)^2\;\sigma^2_\kappa}
\end{equation}

where $\sigma_{m_0}$, $\sigma_C$, $\sigma_\gamma$ and $\sigma_K$ are the
uncertainties on the zero point, sky colour, colour term and extinction
coefficient respectively. Using the proper numbers one can see that the 
typical expected global error is 0.03$\div$0.04 mag, with the measures 
in $V$ and $R$ slightly more accurate than in $U$, $B$ and $I$.

\section{\label{sec:survey} ESO-Paranal night sky brightness survey}

The results we will discuss have been obtained between April 1 2000 
and September 30 2001, corresponding to ESO Observing Periods 65, 66 and 67,
and include data obtained on 174 different nights.
During these eighteen months, 4439 images taken in the $UBVRI$ 
passbands and processed by the FORS pipeline were analysed and 3883 of them
($\sim$88\%) were judged to be suitable for sky brightness measurements, 
according to the criteria we have discussed in Patat (\cite{patat}). 
The numbers for the different passbands are shown in Table~\ref{tab:filt}, 
where we have reported the total number $N_t$ of examined frames, the number 
$N_s$ of frames which passed the tests and the percentage of success $f_s$. 
As expected, this is particularly poor for the $U$ filter, where the sky 
background level is usually very low. In fact, we have shown that practically 
all frames with a sky background level lower than 400 electrons are rejected
(Patat \cite{patat}, Sec.~5).
Since to reach this level in the $U$ passband one needs to expose for more 
than 800 seconds (see Table~\ref{tab:rates}), this explains the large 
fraction of unacceptable frames.
We also note that the number of input frames in the various filters reflects
the effective user's requests. As one can see from Table~\ref{tab:filt}, the
percentage of filter usage $f_t$ steadily grows going from blue to red 
filters, with $R$ and $I$ used in almost 70\% of the cases, while the $U$ 
filter is extremely rarely used.

\begin{table}
\centering
\caption{\label{tab:filt} Number of sky brightness measurements obtained
with FORS1 in $U,B,V,R,I$ passbands from April 1, 2000 to September 30, 2001.}
\begin{tabular}{ccccc}
\hline \hline
 Filter & $f_t$ (\%)  &$N_t$  & $N_s$   & $f_s$ (\%) \\
\hline
U  & 1.8  &  204 & 68   & 33.3 \\
B  & 11.3 &  479 & 434  & 90.6 \\
V  & 17.3 &  845 & 673  & 79.6 \\
R  & 27.1 & 1128 & 1055 & 93.5 \\
I  & 42.5 & 1783 & 1653 & 92.7 \\
\hline
   &      & 4439 & 3883 & 87.5 \\
\hline
\end{tabular}
\end{table}

\begin{figure}
\resizebox{\hsize}{!}{\includegraphics{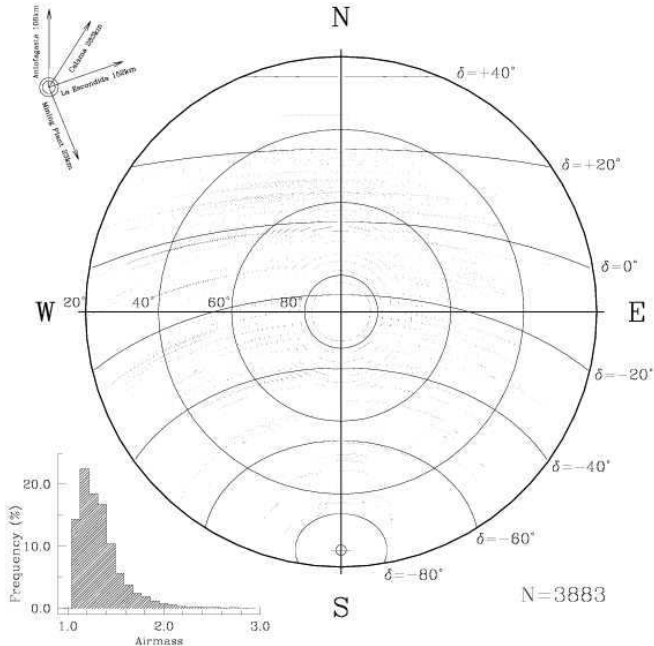}}
\caption{\label{fig:altaz} Distribution of telescope pointings in
Alt-Az coordinates. The lower left insert shows the airmass 
distribution.
}
\end{figure}

%Given these numbers, the present sample of broad band sky brightness 
%measurements is the largest ever published.

To allow for a thorough analysis of the data, the sky brightness measurements
have been logged together with a large set of parameters, some of which are
related to the target's position and others to the ambient conditions.
The first set has been computed using routines adapted from those coded by 
J. Thorstensen\footnote{The original routines by J. Thorstensen are available
at the following ftp server: ftp://iraf.noao.edu/contrib/skycal.tar.Z.} 
and it includes average airmass, azimuth, galactic longitude and latitude, 
ecliptic latitude and helio-ecliptic longitude,
target-moon angular distance, moon elevation, fractional lunar illumination 
(FLI), target-Sun angular distance, Sun elevation and time distance between 
observation and closest twilight. Additionally we have implemented two 
routines to compute the expected moon brightness and zodiacal light 
contribution at target's position.
For the first task we have adopted the model by Krisciunas \& Schaefer 
(\cite{krischae}) and its generalisation to $UBVRI$ passbands (Schaefer 1998), 
while for the zodiacal light we have applied a bi-linear interpolation to 
the data presented by Levasseur-Regourd \& Dumont (\cite{levasseur}). The 
original data is converted from $S_{10}(V)$ to $cgs$ units and the $UBRI$ 
brightness is computed from the $V$ values reported by Levasseur-Regourd \& 
Dumont assuming a solar spectrum, which is a good approximation in the 
wavelength range 0.2$-$2$\mu$m (Leinert et al. 1998). For this purpose 
we have adopted the Sun colours reported by Livingston (\cite{livingston}),
which turn into the following  $U$, $B$, $R$ and $I$ zodiacal light 
intensities normalised to the $V$ passband: 0.52, 0.94, 0.77 and 0.50.
For a derivation of the conversion factor from $S_{10}(V)$ to $cgs$ units,
see Appendix~\ref{sec:units}.

The ambient conditions were retrieved from the VLT Astronomical Site Monitor
(ASM, Sandrock et al. 2000). For our purposes we have included air 
temperature, relative humidity, air pressure, wind speed and wind direction, 
averaging the ASM entries across the exposure time. 
Finally, to allow for further quality selections, for each sky brightness 
entry we have logged the number of sub-windows which passed the $\Delta$-test 
(see Patat \cite{patat}) and the final number $n_g$ of selected sub-windows 
effectively used for the background estimate.

Throughout this paper the sky brightness is expressed in mag arcsec$^{-2}$,
following common astronomical practice. However, when
one is to correct for other effects (like zodiacal light or scattered moon
light), it is more practical to use a linear unit. For this purpose, when
required, we have adopted the $cgs$ system, where the sky brightness is
expressed in erg s$^{-1}$ cm$^{-2}$ \AA$^{-1}$ sr$^{-1}$. In these units the 
typical sky brightness varies in the range 
10$^{-9}-10^{-6}$ (see also Appendix~\ref{sec:units}). It is natural to 
introduce a surface brightness unit (sbu) defined as 
1 sbu$\equiv$10$^{-9}$ erg s$^{-1}$ cm$^{-2}$ \AA$^{-1}$ sr$^{-1}$. In the 
rest of the paper we will use this unit to express the sky brightness in a 
linear scale.

\begin{figure}
\resizebox{\hsize}{!}{\includegraphics{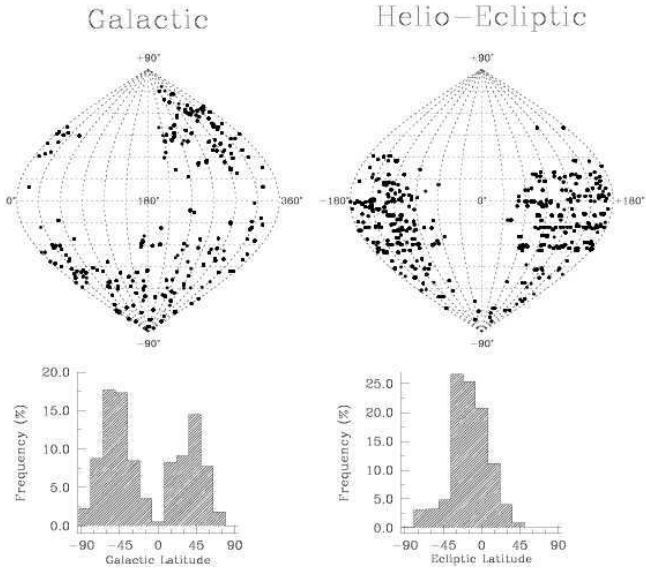}}
\caption{\label{fig:coords} Distribution of telescope pointings in
galactic (left panel) and helio-ecliptic (right panel) coordinates. The
two histograms show the distribution of galactic and ecliptic latitudes.
}
\end{figure}

Due to the large number of measurements, the data give a good coverage of 
many relevant parameters. This is fundamental, if one is to 
investigate possible dependencies. In the next sub-sections we describe
the statistical properties of our data set with respect to these parameters,
whereas their correlations are discussed in 
Secs.~\ref{sec:dark}-\ref{sec:solar}.

\subsection{\label{sec:coord}Coordinate distribution}

The telescope pointings are well distributed in azimuth and elevation, as it 
is shown in Figure~\ref{fig:altaz}. In particular, they span a good range in 
airmass, with a few cases reaching zenith distances larger than 60$^\circ$. 
Due to the Alt-Az mount of the VLT, the region close to zenith is not 
observable, while for safety reasons the telescope does not point at zenith 
distances larger than 70$^\circ$. Apart from these two avoidance areas, the 
Alt-Az space is well sampled, at least down to zenith distances 
$Z$=50$^\circ$. At higher airmasses the western side of the sky appears to 
be better sampled, due to the fact that the targets are sometimes followed 
well after the meridian, while the observations tend to start when
they are on average at higher elevation.

Since sky brightness is expected to depend on the observed position with
respect to the Galaxy and the Ecliptic (see Leinert et al. 1997 for an 
extensive review), it is interesting to see how our measurements are
distributed in these two coordinate systems. Due to the kind of scientific
programmes which are usually carried out with FORS1, we expect that most
of the observations are performed far from the galactic plane. This is
confirmed by the left panel of Figure~\ref{fig:coords}, where we have plotted
the galactic coordinates distribution of the 3883 pointing included in our 
data set. As one can see, the large majority of the points lie at 
$|b|>$10$^\circ$, and therefore the region close to the galactic plane is 
not well enough sampled to allow for a good study of the sky brightness 
behaviour in that area.

\begin{figure}
\resizebox{\hsize}{!}{\includegraphics{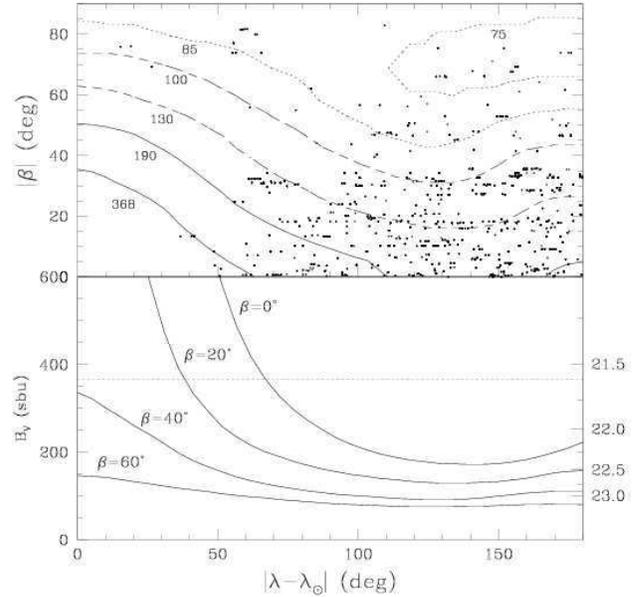}}
\caption{\label{fig:zl} Upper panel: distribution of telescope pointings
in helio-ecliptic coordinates. Over imposed is a contour plot of the
zodiacal light $V$ brightness at the indicated levels expressed in sbu
(1 sbu = 10$^{-9}$ erg s$^{-1}$ cm$^{-2}$ \AA$^{-1}$ sr$^{-1}$). Original data
are from Levasseur-Regourd \& Dumont (\cite{levasseur}). 
Lower panel:  Zodiacal light $V$ brightness profiles at four different 
ecliptic latitudes expressed in sbu (left scale) and mag arcsec$^{-2}$ 
(right scale). The brightness increase seen at
$|\lambda-\lambda_\odot|>$150$^\circ$ is the so-called {\it Gegenschein}.
The horizontal dotted line is placed at a typical $V$ global sky
brightness during dark time (21.6 mag arcsec$^{-2}$). 
}
\end{figure}

The scenario is different if we consider the helio-ecliptic coordinate system
(Figure~\ref{fig:coords}, right panel). The observations are well distributed
across the ecliptic plane for $|\lambda-\lambda_\odot|>$60$^\circ$ and
$\beta<$+30$^\circ$, where the contribution of the zodiacal light to the
global sky brightness can be significant. As a matter of fact, the
large majority of the observations have been carried out in the range
$-$30$^\circ \leq \beta\leq+$30$^\circ$, where the zodiacal light is
rather important at all helio-ecliptic longitudes. This is clearly visible
in the upper panel of Figure~\ref{fig:zl}, where we have over imposed the 
telescope pointings on a contour plot of the zodiacal light $V$ brightness, 
obtained from the data published by Levasseur-Regourd \& Dumont 
(\cite{levasseur}).

We note that the wavelength dependency of the zodiacal light contribution
is significant even within the optical range. In particular it reaches 
its maximum contribution in the $B$ passband,
where the ratio between zodiacal light and typical dark time sky 
flux is always larger than 30\%. On the opposite side we have the $I$ 
passband, where for $|\lambda-\lambda_\odot|\geq$80$^\circ$ the 
contribution is always smaller than 30\% (see also O'Connell 1987).
Now, using the data from Levasseur-Regourd \& Dumont (\cite{levasseur})
and the typical dark time sky brightness measured on Paranal, we can
estimate the sky brightness variations one expects on the basis of the
pure effect of variable zodiacal light contribution. As we have already 
mentioned,
the largest variation is expected in the $B$ band, where already 
at $|\lambda-\lambda_\odot|$=90$^\circ$ the sky becomes inherently brighter
by 0.4-0.5 mag as one goes from $|\beta|>$60$^\circ$ to $|\beta|=$0$^\circ$.
This variation decreases to $\sim$0.15 mag in the $I$ passband.

Due to the fact that the bulk of our data has been obtained at
$|\beta|\leq$30$^\circ$, our dark time sky brightness estimates are 
expected to be affected by systematic zodiacal light effects, which have
to be taken into account when comparing our results with those obtained
at high ecliptic latitudes for other astronomical sites (see 
Sec.~\ref{sec:dark}).

\subsection{\label{sec:moon}Moon Contribution}

Another relevant aspect that one has to take into account when measuring the 
night sky brightness is the contribution produced by scattered moon light.
Due to the scientific projects FORS1 was designed for, the large majority of 
observations are carried out in dark time, either when the fractional lunar 
illumination (FLI) is small or when the moon is below the horizon. 
Nevertheless, according to the user's requirements, some observations are 
performed when moon's contribution to the sky background is not negligible. 
To evaluate the amount of moon light contamination at a given position on the 
sky (which depends on several parameters, like target and moon elevation, 
angular distance, FLI and extinction coefficient in the given passband) we 
have used the model developed by Krisciunas and Schaefer (\cite{krischae}), 
with the double aim of selecting those measurements which are not influenced 
by moonlight and to test the model itself. We have forced
the lunar contribution to be zero when moon elevation is 
$h_m\leq-$18$^\circ$ and we have neglected any twilight effects.
On the one hand this has certainly the effect of overestimating the moon 
contribution when $-$18$^\circ \leq h_m \leq$0$^\circ$, but on the other 
hand it puts us on the safe side when selecting dark time data. 

As expected, a large fraction of the observations were obtained 
practically with no moon: in more than 50\% of the cases moon's addition is 
from 10$^{-1}$ to 10$^{-3}$ the typical dark sky brightness. Nevertheless, 
there is a substantial tail of observations where the contamination is 
relevant (200$-$400 sbu) and a few extreme cases were the moon is the 
dominating source ($>$600 sbu). This offers us the possibility of exploring 
both regimes.

\subsection{\label{sec:sunactiv} Solar activity}

\begin{figure}
\resizebox{\hsize}{!}{\includegraphics{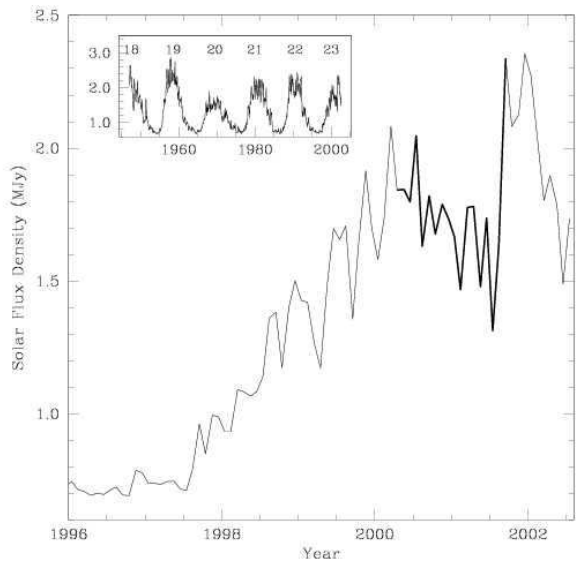}}
\caption{\label{fig:sun} Penticton-Ottawa Solar flux at 2800 MHz (monthly 
average). The time range covered by the data presented in this paper is 
indicated by the thick line. The upper left insert traces the solar flux 
during the last six cycles.
}
\end{figure}

We conclude the description of the statistical properties of our data set 
by considering the solar activity during the relevant time interval. 
As it has been first pointed out by Lord Rayleigh (\cite{rayleigh}), the 
airglow emission is correlated with the sunspot number. As we have seen in
Sec.~\ref{sec:intro}, this has been confirmed by a number of studies and 
is now a widely accepted effect (see Walker 1988b and references therein). 
As a matter of fact, all measurements presented in this work were taken 
very close to the maximum of sunspot cycle n.~23, and thus we do not expect 
to see any clear trend. This is shown in Figure~\ref{fig:sun}, where 
we have plotted the monthly averaged Penticton-Ottawa solar flux at 2800 MHz 
(Covington 1969)\footnote{The data are available in digital form at the 
following web site: {\tt http://www.drao.nrc.ca/icarus/www/archive.html}.}. 
We notice that the solar flux abruptly changed by a factor $\sim$2 
between July and September 2001, leading to a second maximum which lasted
roughly two months at the end of year 2001. This might have some effect on 
our data, which we will discuss later on.

\section{\label{sec:dark} Dark time sky brightness at Paranal}

\begin{figure}
\resizebox{\hsize}{!}{\includegraphics{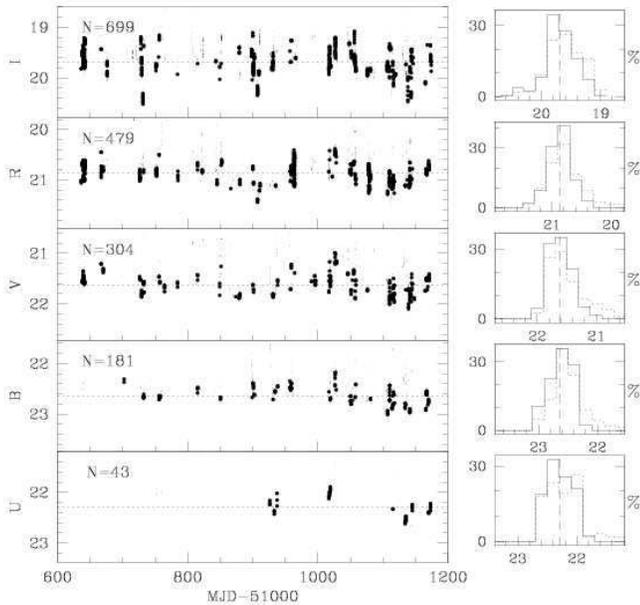}}
\caption{\label{fig:dark} Zenith corrected sky brightness measured at Paranal
during dark time (thick dots) from April 1st, 2000 to September 30th, 2001.
The selection criteria are: 
$|b|>$10$^\circ$, $|\lambda-\lambda_\odot|\geq$90$^\circ$, 
$\Delta t_{twi}>$1 hour, FLI=0 or $h_m\leq-$18$^\circ$.
Thin dots indicate all observations (corrected to zenith). The 
horizontal dotted lines are positioned at the average values of the
selected points.
The histograms trace the distribution of selected measurements (solid
line) and all measurements (dotted line), while the vertical dashed lines
are placed at the average sky brightness during dark time.
}
\end{figure}

Due to the fact that our data set collects observations performed under
a wide range of conditions, in order to estimate the zenith sky brightness 
during dark time it is necessary to apply some selection. For this
purpose we have adopted the following criteria: photometric conditions, 
airmass $X\leq$1.4, galactic latitude $|b|>$10$^\circ$, time distance from 
the closest twilight $\Delta t_{twi}>$1 hour and no moon (FLI=0 or 
$h_M\leq-$18$^\circ$). Unfortunately,
as we have mentioned in Sec.~\ref{sec:coord}, very few observations have
been carried out at $|\beta|>$45$^\circ$ and hence we could not put
a very stringent constraint on the ecliptic latitude, contrary to what is 
usually done (see for example Benn \& Ellison 1998). To limit the 
contribution of the zodiacal light, we could only restrict the range of 
helio-ecliptic longitude ($|\lambda-\lambda_\odot|\geq$90$^\circ$).  
The results one obtains from this selection are summarized in 
Table~\ref{tab:dark1} and Figure~\ref{fig:dark}, where we have plotted the 
estimates of the sky brightness at zenith as a function of time. 
Once one has accounted for the zodiacal light
bias (see below), the values are consistent with those
reported for other dark sites; in particular, they are very similar to
those presented by Mattila et al. (\cite{attila}) for La Silla, which were
also obtained during a sunspot maximum (February 1978). 
As pointed out by several authors, the dark time values show quite a strong 
dispersion, which is typically of the order of 0.2 mag RMS. Peak to peak 
variations in the $V$ band are as large as 0.8 mag, while this excursion 
reaches 1.5 mag in the $I$ band.

\begin{table}
\centering
\caption{\label{tab:dark1} Zenith corrected average sky brightness during 
dark time at Paranal. Values are expressed in mag arcsec$^{-2}$.
Columns 3 to 7 show the RMS deviation, minimum and maximum brightness,
number of data points and expected average contribution from the zodiacal
light, respectively.}
\begin{tabular}{ccccccccc}
\hline \hline
Filter & Sky Br. & $\sigma$ & Min & Max & N & $\Delta m_{ZL}$\\
\hline
U & 22.28 & 0.22 & 21.89 & 22.61  &  39 &0.18 \\
B & 22.64 & 0.18 & 22.19 & 23.02  & 180 &0.28 \\ 
V & 21.61 & 0.20 & 20.99 & 22.10  & 296 &0.18 \\
R & 20.87 & 0.19 & 20.38 & 21.45  & 463 &0.16 \\
I & 19.71 & 0.25 & 19.08 & 20.53  & 580 &0.07 \\
\hline 
\end{tabular}
\end{table}

In Sec.~\ref{sec:coord} we have shown that the estimates presented in
Table~\ref{tab:dark1} are surely influenced by zodiacal light effects of 
low ecliptic latitudes. To give an idea of the amplitude of this
bias, in the last column of Table~\ref{tab:dark1} we have reported the 
correction $\Delta m_{ZL}$ one would have to apply to the average values 
to compensate for this contribution. This has been computed as the average 
correction derived from the data of Levasseur-Regourd \& Dumont 
(\cite{levasseur}), assuming typical values for the dark time sky brightness: 
as one can see, $\Delta m_{ZL}$ is as large as $\sim$0.3 mag in the 
$B$ passband.

\begin{figure}
\resizebox{\hsize}{!}{\includegraphics{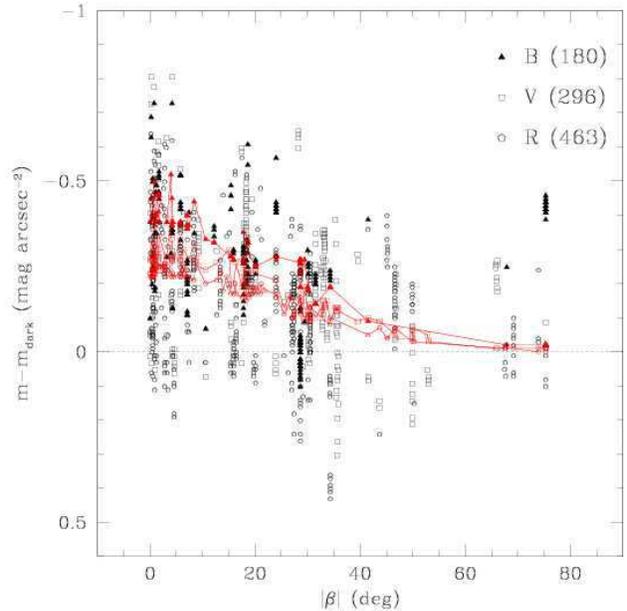}}
\caption{\label{fig:zlplot} $B, V$ and $R$ dark time sky brightness variations
as a function of ecliptic latitude. The solid lines trace the behaviour 
expected from Levasseur-Regourd \& Dumont (\cite{levasseur}) data for the 
different passbands.
}
\end{figure}

The sky brightness dependency on the ecliptic latitude is clearly displayed
in Figure~\ref{fig:zlplot}, where we have plotted the deviations from the 
average sky brightness (cf. Table~\ref{tab:dark1}) for $B, V$ and $R$ 
passbands, after applying the correction $\Delta m_{ZL}$. We have 
excluded the $I$ band because it is heavily dominated by airglow variations, 
which completely mask any dependency from the position in the helio-ecliptic 
coordinate system; the $U$ data were also not included due to the small 
sample. For comparison, in the same figure we have over imposed 
the behaviour expected on the basis of Levasseur-Regourd \& Dumont 
(\cite{levasseur}) data, which have been linearly interpolated to each of the 
positions $(\lambda-\lambda_\odot,\beta)$ in the data set.
As one can see, there is a rough agreement, the overall spread being quite 
large. This is visible also in a similar plot produced by Benn \& 
Ellison (1998, their Figure~10) and it is probably due to the night-to-night
fluctuations in the airglow contribution.

\section{\label{sec:variations} Sky brightness variations during the night}

In the literature one can find several and discordant results about the
sky brightness variations as a function of the time distance from
astronomical twilight. Walker (\cite{walker88b}) first pointed out that the
sky at zenith gets darker by $\sim$0.4 mag arcsec$^{-2}$ during the 
first six hours after the end of twilight. Pilachowski et al. (\cite{pila}) 
found dramatic short time scale variations, while the steady 
variations were attributed to airmass effects only (see their Figure~2). This
explanation looks indeed reasonable, since the observed sky 
brightening is in agreement with the predictions of Garstang 
(\cite{garstang89}).
Krisciunas (1990, his Figure~6) found that his data obtained in the
$V$ passband showed a decrease of $\sim$0.3 mag arcsec$^{-2}$ in the
first six hours after the end of twilight, but he also remarked that this 
effect was not clearly seen in $B$.
Due to the high artificial light pollution, Lockwood, Floyd \& Thompson 
(\cite{lockwood}) tend to attribute the nightly sky brightness decline they
observe at the Lowell Observatory, to progressive reduction of 
commercial activity. 

Walker's findings were questioned by Leinert et al. 
(\cite{leinert95}) and Mattila et al. (\cite{attila}), who state that 
{\it no indications for systematic every-night behaviour of a decreasing sky 
brightness after the end of twilight} were shown by their observations. 
Krisciunas (\cite{krisc97}) notes that, on average, the zenith sky 
brightness over Mauna Kea shows a {\it not very convincing} sky brightness 
change of 0.03 mag hour$^{-1}$. On the other hand he also reports cases where 
the darkening rate was as large as 0.24 mag hour$^{-1}$ and discusses the 
possibility of a reverse Walker effect taking place during a few hours 
before the beginning of morning twilight.

Leinert et al. (\cite{leinert98}) touch this topic in their extensive 
review, pointing out that this is an often observed effect due to a
decreasing release rate of the energy stored in the atmospheric 
layers during day time. Finally, Benn \& Ellison (\cite{benn}) do not find 
any signature of steady sky brightness variation depending on the time 
distance from twilights at La Palma, and suggest that the effect observed by 
Walker is due to the variable contribution of the zodiacal light, a
hypothesis already discussed by Garstang (\cite{garstang97}). A further
revision of Walker's findings is presented here in Appendix \ref{sec:walker},
where we show that the effect is significantly milder than it was thought and
probably influenced by a small number of well sampled nights.

\begin{figure}
\resizebox{\hsize}{!}{\includegraphics{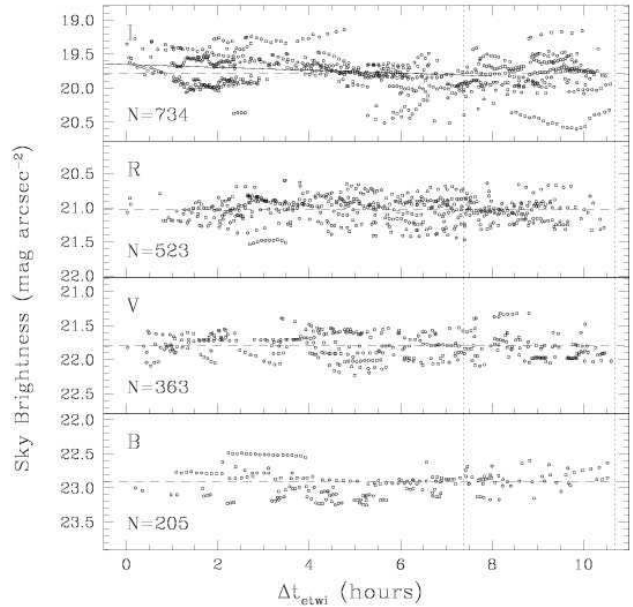}}
\caption{\label{fig:dtwitrend} Dark time Paranal night sky brightness,
corrected for zodiacal light contribution, as a function of time distance 
from evening twilight. The vertical dotted lines indicate the shortest and 
longest night (7.4 and 10.7 hours respectively, astronomical twilight to 
twilight), while the dashed horizontal line is placed at the average value
in each passband.}
\end{figure}

\begin{figure}
\resizebox{\hsize}{!}{\includegraphics{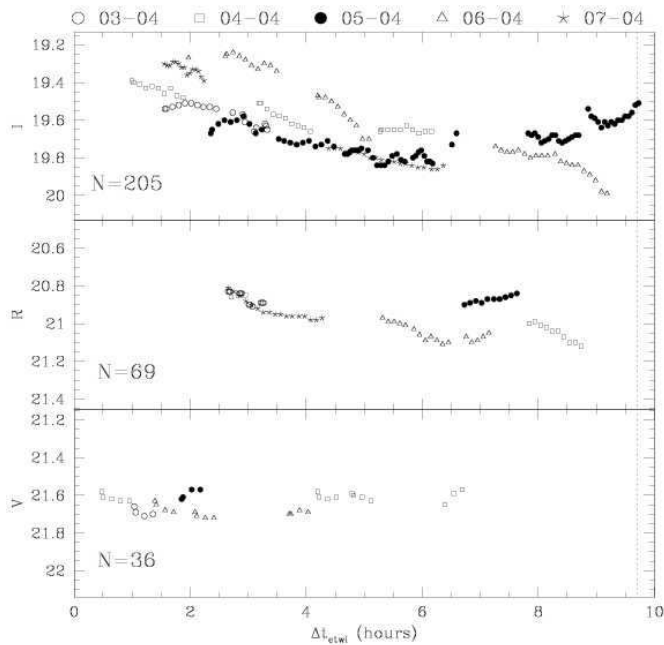}}
\caption{\label{fig:examples3} Time sequences collected on April 2--7, 2000.
The data have been corrected for airmass and differential zodiacal light
contribution. The vertical dotted line is placed at the beginning of morning
astronomical twilight.}
\end{figure}

We have performed an analogous analysis on our
data set, using only the measurements obtained during dark time and
correcting for differential zodiacal light contribution. Since the time
range covered by our observations is relatively small with respect to
the solar cycle, we do not expect the solar activity to play a relevant
role, and hence we reckon it is reasonable not to normalise the measured
sky brightness to some reference time. This operation would be anyway very
difficult, due to the vast amount of data and the lack of long time
series. The results are presented in Figure~\ref{fig:dtwitrend}, where we 
have plotted the sky brightness vs. time from evening twilight,
$\Delta t_{etwi}$, for $B, V, R$ and 
$I$ passbands. Our data do not support the exponential drop seen by
Walker (\cite{walker88b}) during the first 4 hours and confirm the findings
by Leinert et al. (\cite{leinert95}), Mattila et al. (\cite{attila}) and 
Benn \& Ellison (\cite{benn}). This is particularly true for $V$ and $R$ data,
while in $B$ and especially in $I$ one might argue that some evidence of
a rough trend is visible. As a matter of fact, a blind linear least squares
fit in the range 0 $\leq \Delta t_{etwi}\leq$ 6 gives an average slope 
of 0.04$\pm$0.01 and 0.03$\pm$0.01 mag hour$^{-1}$ for the two passbands 
respectively. Both values are a factor of two smaller than those
found by Walker (\cite{walker88b}) but are consistent, within the quoted 
errors, with the values we found revising his original data
(see Appendix \ref{sec:walker}). 
However, the fact that no average steady decline is seen in $V$ and $R$ casts
some doubt on the statistical significance of the results one gets from
$B$ and $I$ data.
This does not mean that on some nights very strong declines can be seen, as
already pointed out by Krisciunas (\cite{krisc97}). Our data set includes 
several such examples, but probably the most interesting is the one which is
shown in Figure~\ref{fig:examples3}, where we have plotted the data collected
on five consecutive nights (2000 April 3--7). As one can see, the $I$ data
(upper panel) show a clear common trend, even though segments with different 
slopes are present and the behaviour shown towards the end of the night during
2000 April 6 is opposite to that of 2000 April 5. This trend becomes less 
clear in the $R$ passband (middle panel) and it is definitely not visible in 
$V$ (lower panel), where the sky brightness remains practically constant for 
about 6 hours. Unfortunately, no $B$ data are available during these nights.

\begin{figure}
\resizebox{\hsize}{!}{\includegraphics{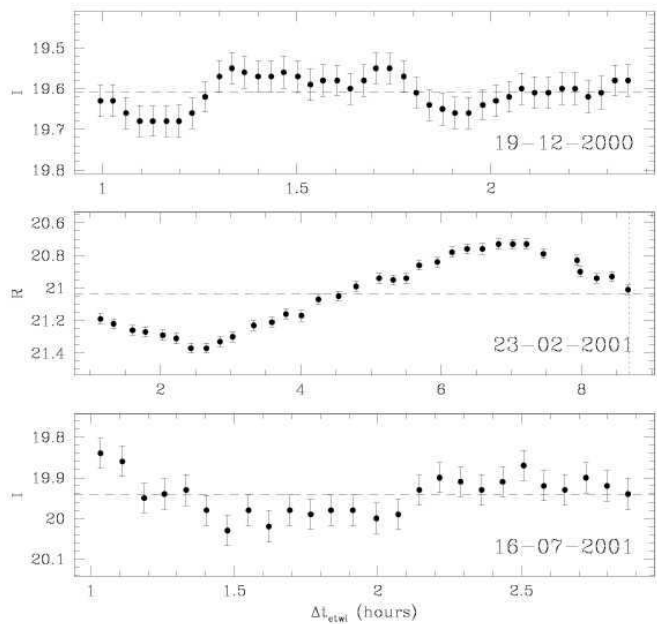}}
\caption{\label{fig:examples4} Time sequences collected on 19-12-2000 ($I$),
23-02-2001 ($R$) and 16-07-2001 ($I$). The data have been corrected for 
airmass and differential zodiacal light contribution. The vertical dotted 
line is placed at the beginning of morning astronomical twilight.}
\end{figure}

A couple of counter-examples are shown in Figure~\ref{fig:examples4}: the
upper and lower panels show two well sampled time series obtained on the
same sky patch, which show that during those nights the sky brightness
was roughly constant during the phase where the Walker effect is expected
to be most efficiently at work. Instead of a steady decline, clear and smooth 
sinusoidal fluctuations with maximum amplitudes of $\sim$0.1 mag and time 
scales of the order of 0.5 hours are well visible.  Finally, to show that 
even mixed behaviours can take place, in the central panel we have presented 
the $R$ data collected on 23-02-2001, when a number of different sky patches 
was observed. During that night, the sky  brightness had a peak-to-peak
fluctuation of $\sim$0.7 mag and showed a steady increase for at least
4 hours.

To conclude, we must say that we tend to agree with Leinert et al. 
(\cite{leinert95}) that the behaviour shown during single 
nights covers a wide variety of cases and that there is no clear average 
trend. We also add that mild time-dependent effects cannot be ruled out; 
they are probably masked by the much wider night-to-night fluctuations and 
possibly by the patchy nature of the night sky even during the same night. 

\section{\label{sec:moonmod} Testing the moon brightness model}

As we have mentioned in Sec.~\ref{sec:moon}, some data have been collected
when the moon contribution to the sky brightness is conspicuous and this
offers us the possibility of directly measuring its effect and comparing it
with the model by Krisciunas \& Schaefer (\cite{krischae}) which, to our
knowledge, is the only one available in the literature.

\begin{figure}
\resizebox{\hsize}{!}{\includegraphics{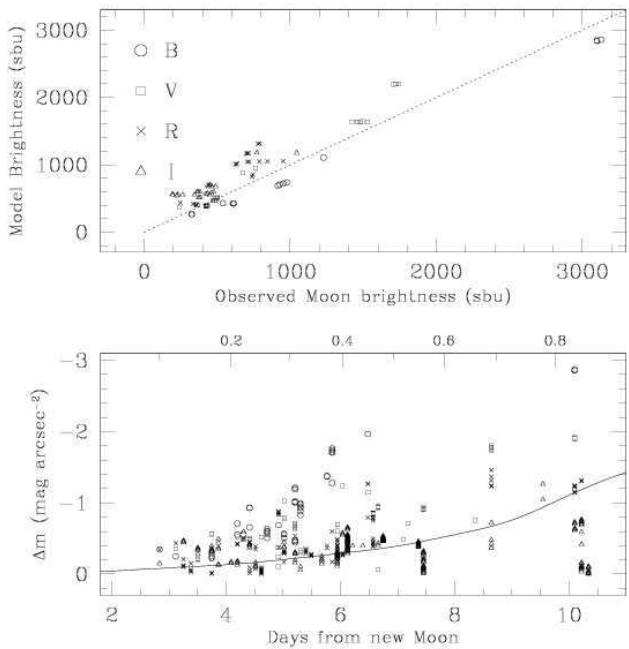}}
\caption{\label{fig:modeltest} Lower panel: observed sky brightness variation
as a function of moon age for $B,V,R$ and $I$. The solid line traces the data
published by Walker (\cite{walker87}) for the $V$ passband while the upper 
scale shows the fractional lunar illumination. Upper panel:
comparison between the observed and predicted moon contribution
(Krisciunas \& Schaefer \cite{krischae}).
Plotted are only those data points for which the global brightness is larger
than the typical dark time brightness. 
}
\end{figure}

To estimate the fraction of sky brightness generated by scattered moon
light, we have subtracted to the observed fluxes the average values 
reported in Table~\ref{tab:dark1} for each passband. The results are presented
in the lower panel of Figure~\ref{fig:modeltest}, where we have plotted only 
those data for which the observed value was larger than the dark time one.
As expected, the largest deviations are seen in $B$, where the sky brightness
can increase by about 3 mag at 10 days after new moon, while in $I$, at 
roughly the same moon age, this deviation just reaches $\sim$1.2 mag. It is
interesting to note that most exposure time calculators for modern instruments
make use of the function published by Walker (\cite{walker87}) to compute
the expected sky brightness as a function of moon age. As already noticed
by Krisciunas (\cite{krisc90}), this gives rather optimistic estimates, real
data being most of the time noticeably brighter. This is clearly visible in
Figure~\ref{fig:modeltest}, where we have overplotted Walker's function
for the $V$ passband to our data: already at 6 days past new moon the 
observed $V$ data (open squares) show maximum deviations of the order of 1 
mag. These results are fully compatible with those presented by Krisciunas 
(\cite{krisc90}) in his Figure~8. 

Another weak point of Walker's function is that it has one input parameter
only, namely the moon phase, and this is clearly not enough to predict
with sufficient accuracy the sky brightness. This, in fact, depends on
a number of parameters, some of which, of course, are known only when the
time the target is going to be observed is known. 
In this respect, the model by Krisciunas \& Schaefer  (\cite{krischae}) is
much more promising, since it takes into account all relevant astronomical
circumstances. The model accuracy was tested by the authors themselves,
who reported RMS deviations as large as 23\% in a brightness range which
spans over 20 times the typical value observed during dark time.

In the upper panel of Figure~\ref{fig:modeltest} we have compared our results 
with the model predictions, including $B,V,R$ and $I$ data. We emphasise 
that we have used average values for the extinction coefficients and dark 
time sky brightness and this certainly has some impact on the computed 
values. On the other hand, this is the typical configuration 
under which the procedure would be implemented in an exposure time 
calculator, and hence it gives a realistic evaluation of the model 
practical accuracy. Figure~\ref{fig:modeltest} shows that, even if deviations
as large as 0.4 mag are detected, the model gives a reasonable reproduction 
of the data in the brightness range covered by our observations. This is
actually less than half with respect to the one encompassed by the data 
shown in Figure~3 of Krisciunas \& Schaefer (\cite{krischae}), which reach 
$\sim$8300 sbu in the $V$ band.

\section{\label{sec:solar} Sky brightness vs. solar activity}

\begin{figure}
\resizebox{\hsize}{!}{\includegraphics{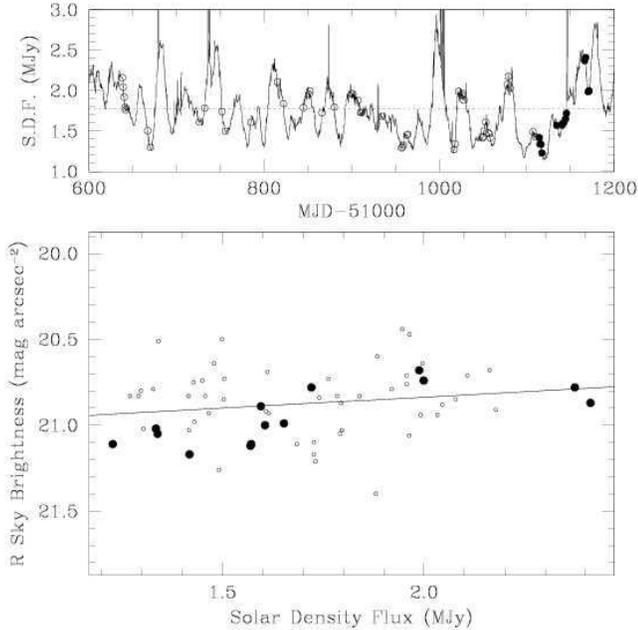}}
\caption{\label{fig:sunave} Lower panel: nightly average sky brightness in
the $R$ passband vs. solar density flux. The filled circles represent
data taken at MJD$>$52114, while the solid line indicates a linear least
squares fit to all data. Upper panel: Penticton-Ottawa solar flux at 
2800 MHz during the time interval discussed in this paper. The open 
circles indicate the values which correspond to the data presented in 
the lower panel and the dotted line is placed at the median value for the
solar density flux.
}
\end{figure}

As we have mentioned in Sec.~\ref{sec:sunactiv}, during the time covered
by the data presented here, the solar activity had probably reached its
maximum. To be more precise, since the current solar cycle (n.~23) has
a double peak structure (see Figure~\ref{fig:sun}), our measurements cover
the descent from the first maximum and the abrupt increase to the second
maximum. Mainly due to the latter transition, the solar density flux 
at 10.7 cm in our data set ranges from 1.2 MJy to 2.4 MJy, the median value 
being 1.8 MJy.
Even though this is almost half of the full range expected on a typical 
complete 11 years solar cycle (80$-$250 MJy), a clear
variation is seen in the same solar density flux range from similar analysis
performed by other authors (see for example Mattila et al. 1996, their
Figure~6). In Figure~\ref{fig:sunave} we show the case of the $R$ passband, where
we have plotted the nightly average sky brightness vs. the solar density flux
measured during the day immediately preceeding the observations. A linear
least squares fit to the data (solid line) gives a slope of 
0.14$\pm$0.01 mag arcsec$^{-2}$ MJy$^{-1}$, which turns
into a variation of 0.24$\pm$0.11 mag arcsec$^{-2}$ during a full solar 
cycle. This value is a factor two smaller than what has been reported
for $B$, $V$ (Walker 1988b, Krisciunas 1990) and $uvgyr$ 
(Leinert et al. 1995, Mattila et al. 1996) for yearly averages and it is
consistent with a null variation at the 2 sigma level. Moreover,
since the correlation factor computed for the data in Figure~\ref{fig:sunave}
is only 0.19, we think there is no clear indication for a real dependency.

This impression is confirmed by the fact that a similar analysis for
the $B$ and $V$ passbands gives an extrapolated variation of 0.08$\pm$0.13 
and 0.07$\pm$0.11 mag arcsec$^{-2}$ respectively. These numbers, which
are consistent with zero, and the low correlation coefficients
(0.08 and 0.11 respectively) seem to indicate no short--term dependency
from the 10.7 cm solar flux. Similar values
are found for the $I$ passband ($\Delta m$=0.22$\pm$0.15 mag arcsec$^{-2}$).
These results agree with the findings by Leinert et al. (\cite{leinert95}) 
and Mattila et al. (\cite{attila}) and the early work of 
Rosenberg \& Zimmermann (\cite{rosenberg}), who have shown that the 
[OI]5577\AA\/ line intensity correlates with the 2800MHz solar flux much more 
strongly using the monthly averages than the nightly averages.
For all these reasons, we agree with Mattila et al. (\cite{attila}) in saying 
that no firm prediction on the night sky brightness can be made on the basis
of the solar flux measured during the day preceeding the observations,
as it was initially suggested by Walker (\cite{walker88b}).
A possible physical explanation for this effect is that there is some inertia
in the energy release from the layers ionised by the solar UV radiation, such 
that what counts is the integral over some typical time scale rather than 
the instantaneous energy input.

\section{\label{sec:discuss} Discussion and conclusions}

\begin{table*}
\centering
\caption{\label{tab:compare} Dark time zenith night sky brightness measured
at various observatories (adapted from Benn \& Ellison 1998). $S_{10.7cm}$ 
is the Penticton-Ottawa solar density flux at 2800 MHz (Covington 1969).}
\begin{tabular}{llccccccl}
\hline \hline
Site        & Year   & $S_{10.7cm}$ & $U$ & $B$ & $V$ & $R$ & $I$ & Reference \\
\cline{4-8} &        & MJy          & \multicolumn{5}{c}{mag arcsec$^{-2}$} & \\
\hline
La Silla    & 1978   & 1.5 & -     & 22.8  & 21.7  & 20.8  & 19.5 & Mattila et al. (1996)\\
Kitt Peak   & 1987   & 0.9 & -     & 22.9  & 21.9  & -     & -    & Pilachowski et al. (1989)\\
Cerro Tololo& 1987-8 & 0.9 & 22.0  & 22.7  & 21.8  & 20.9  & 19.9 & Walker (1987, 1988a) \\
Calar Alto  & 1990   & 2.0 & 22.2  & 22.6  & 21.5  & 20.6  & 18.7 & Leinert et al. (1995)\\
La Palma    & 1994-6 & 0.8 & 22.0  & 22.7  & 21.9  & 21.0  & 20.0 & Benn \& Ellison (1998)\\
Mauna Kea   & 1995-6 & 0.8 & -     & 22.8  & 21.9  & -     & -    & Krisciunas (1997)\\
Paranal     & 2000-1 & 1.8 & 22.3  & 22.6  & 21.6  & 20.9  & 19.7 & this work \\
\hline 
\end{tabular}
\end{table*}

Besides being the first systematic campaign of night sky brightness 
measurements at Cerro Paranal, the survey we have presented here has many 
properties that make it rather unique. First of all, the fact that it is 
completely automatic ensures that each single frame which passes through 
the quality checks contributes to build a continously growing sample. Furthermore, 
since the data are produced by a very large telescope, the measurements 
accuracy is quite high when compared to that generally achieved
in this kind of study, which most of the time make use of small telescopes.
Another important fact, related to both the large collecting area and the
use of a CCD detector, is that the usual problem of faint
unresolved stars is practically absent.  In fact, with small telescopes,
it is very difficult to avoid the inclusion of stars fainter than $V$=13
in the beam of the photoelectric photometer (see for example Walker 1988b).
The contribution of such stars is 39.1 $S_{10}(V)$ (Roach \& Gordon \cite{roachgordon},
Table~2-I) which corresponds to about 13\% of the global sky brightness. Now, 
with the standard configuration and a seeing of 1$^{\prime\prime}$, 
during dark time FORS1 can reach a 5$\sigma$ peak limiting magnitude 
$V\simeq$23.3 in a 60 seconds exposure for unresolved objects. As the 
simulations show (see Patat \cite{patat}), the algorithm we have adopted to 
estimate the sky background is practically undisturbed by the presence of 
such stars, unless their number is very large, a case which would be
rejected anyway by the $\Delta$-test (Patat \cite{patat}).
Now, since the typical contribution of stars with $V\geq$20 is 3.2 $S_{10}(V)$
(Roach \& Gordon \cite{roachgordon}), we can conclude that the effect of 
faint unresolved stars on our measurements is less than 1\%.

Another distinguishing feature is the time coverage. As reported by Benn \& 
Ellison (\cite{benn}), the large majority of published sky brightness 
measurements were carried out during a limited number of nights (see their 
Table~1). The only remarkable exception is represented by their own work, 
which made use of 427 CCD images collected on 63 nights in ten years.  
Nevertheless, this has to be compared with our survey which produced about
3900 measurements during the first 18 months of steady operation. This high 
time frequency allows one to carry out a detailed analysis of time dependent 
effects, as we have shown in Sec.~\ref{sec:variations} and to get
statistically robust estimates of the typical dark time zenith sky
brightness. 

\begin{figure}
\resizebox{\hsize}{!}{\includegraphics{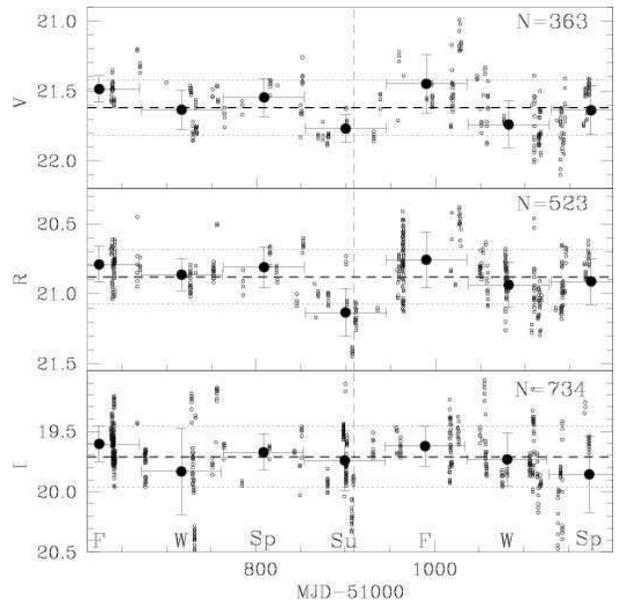}}
\caption{\label{fig:sodium} $V$, $R$ and $I$ (from top to bottom) dark time 
sky brightness measured at Paranal from April 2000 to September 2001.
For each passband the average value (dashed line) and the $\pm$1$\sigma$ 
interval (dotted lines) are plotted. The large solid dots are placed
at the seasonal average values computed within the time interval indicated by
the horizontal error bars, while the vertical error bars represent the 
corresponding RMS sky brightness dispersion.
The labels on the lower side indicate the austral astronomical season: winter
(W), spring (Sp), summer (Su) and fall (F). The vertical dashed line indicates
December 31, 2000.
}
\end{figure}

The values we have obtained for Paranal are compared to
those of other dark astronomical sites in Table~\ref{tab:compare}.
The first thing one notices is that the values for Cerro Paranal are very
similar to those reported for La Silla, which were also obtained during
a maximum of solar activity. They are also not very different from those
of Calar Alto, obtained in a similar solar cycle phase, even though 
Paranal and La Silla are clearly darker in $R$ and definitely in $I$.
All other sites presented in Table~\ref{tab:compare} have data which were
obtained during solar minima and are therefore expected to show systematically
lower sky brightness values. This is indeed the case. For example, the
$V$ values measured at Paranal are about 0.3 mag brighter then those
obtained at other sites at minimum solar activity (Kitt Peak, Cerro
Tololo, La Palma and Mauna Kea). The same behaviour, even though somewhat
less pronounced, is seen in $B$ and $I$, while it is much less obvious in $R$.
Finally, the $U$ data show an inverse trend, in the sense that at those
wavelengths the sky appears to be brighter at solar minima. Interestingly,
a plot similar to that of Figure~\ref{fig:sunave} also gives 
a negative slope, which turns into a variation $\Delta U$=$-$0.7$\pm$0.5 mag 
arcsec$^{-2}$ during a full solar cycle. Due to the rather large error
and the small number of nights (11), we think that no firm conclusion can be
drawn about a possible systematic effect, but we notice that a similar behaviour is
found by Leinert et al. (\cite{leinert95}) for the $u$ passband (see their
Figure~6). Since the airglow in $U$ is dominated by the O$_2$ Herzberg bands
$A^3\Sigma-X^3\Sigma$ (Broadfoot \& Kendall 1968), the fact that their
intensity seems to decrease with an increasing ionising solar flux could
probably give some information on the physical state of the emitting layers,
where molecular oxygen is confined.

At any rate, the $BVRI$ Paranal sky brightness will probably decrease in the 
next 5-6 years, to reach its natural minimum around 2007. The expected 
darkening is of the order of 0.4-0.5 mag arcsec$^{-2}$ (Walker 1988b), but 
the direct measurements will give the exact values for this particular site. 
In the next years this survey will provide an unprecedented mapping of the 
dependency from solar activity. So far, in fact, this correlation has been 
investigated with sparse data, affected by a rather high spread due to the 
night-to-night variations of the airglow (see for instance Figure~4 by 
Krisciunas 1990), which tend to mask any other effect and make any 
conclusion rather uncertain.

As already pointed out by several authors, the night sky can vary 
significantly over different time scales, following physical processes that 
are not completely understood. As we have shown in the previous section, 
even the daily variations in the solar ionising radiation are not sufficient 
to account for the observed night-to-night fluctuations. Moreover, the 
observed scatter in the dark time sky brightness (see Sec.~\ref{sec:dark}) 
is certainly not produced by the measurement accuracy and can be as large as 
0.25 mag (RMS) in the $I$ passband; since the observed distribution is 
practically Gaussian (see Figure~\ref{fig:dark}), this means that the $I$ sky 
brightness can range over $\sim$1.4 mag, even after removing the effects of 
airmass and zodiacal light contribution. This unpredictable variation has 
the unpleasant effect of causing maximum signal-to-noise changes of about a 
factor of 2.

Besides these short time scale fluctuations that we have discussed in 
Sec.~\ref{sec:variations} and the long term variation due to the solar cycle,
one can reasonably expect some effects on intermediate time scales. With this
respect we have computed the sky brightness values averaged over three months
intervals, centered on solstices and equinoxes. The results for $V$, $R$ and 
$I$ are plotted in Figure~\ref{fig:sodium}, where we have used all the 
available data obtained at Paranal during dark time, with $\Delta_{twi}\geq$0.
This figure shows that there is no convincing 
evidence for any seasonal effect, especially in the $I$ passband, where all 
three-monthly 
values are fully consistent with the global average (thick dashed line). The
only marginal detection of a deviation from the overall trend is that seen in 
$R$ in correspondence of the austral summer of year 2000, when the average 
sky brightness turns out to be $\sim$1.3$\sigma$ fainter than the global 
average value. Even though a decrease of about 0.1 mag is indeed expected in 
the $R$ passband as a consequence of the NaI~D flux variation (see Roach \& 
Gordon \cite{roachgordon} and the discussion below), we are not completely sure this is 
the real cause of the observed effect, both because of the low statistical 
significance and the fact that a similar, even though less pronounced drop, 
is seen at the same epoch in the $V$ band, where the NaI~D line contribution 
is negligible (see Figure~\ref{fig:skyspectra}).

\begin{figure}
\resizebox{\hsize}{!}{\includegraphics{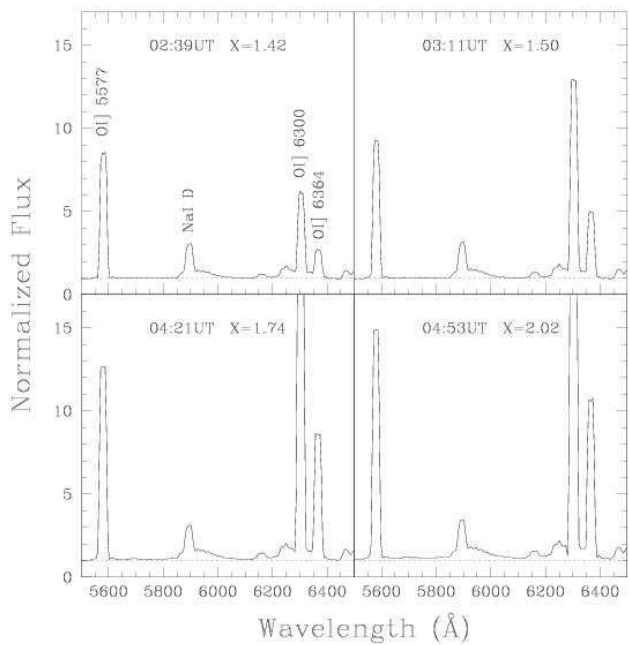}}
\caption{\label{fig:skyevol} Evolution of the night sky spectrum on 
February 25, 2001 in the wavelength range 5500-6500 \AA. The original
1800 seconds spectra were obtained with FORS1, using the standard resolution
collimator and a long slit 1$^{\prime\prime}$ wide (see also the caption of
Figure~\ref{fig:skyspectra}). In each panel the starting 
UT time and airmass $X$ are reported. For presentation the four spectra have been
normalised to the continuum of the first one in the region 5600-5800 \AA.
}
\end{figure}

To illustrate how complex the night sky variations can be, we present a 
sequence of four spectra taken at Paranal during a moonless night in 
Figure~\ref{fig:skyevol}, starting more than two hours after evening
twilight with an airmass ranging from 1.4 to 2.0. For the sake of simplicity
we concentrate on the spectral region 5500-6500\AA, right at the intersection
between $V$ and $R$ passbands, which contains the brightest optical emission 
lines and the so called pseudo-continuum (see Sec.~\ref{sec:intro}). Due to 
the increasing airmass, the overall sky brightness is expected to grow
according to Eq.~\ref{eq:airmdep}, which for $V$ and $R$ gives a variation of
about 0.2 mag. These values are in rough agreement with those
one gets measuring the continuum variation at 5500\AA\/ (0.13 mag) and 
6400\AA\/ (0.18 mag). Interestingly, this is not the case for the
synthetic $V$ and $R$ magnitudes derived from the same spectra, which decrease
by 0.32 and 0.51 mag respectively, i.e. much more than expected, specially
in the $R$ band. This already tells us that the continuum and the emission lines
must behave in a different manner.
In fact, the flux carried by the [OI]5577\AA\/ line changes by a factor 1.9 
from the first to the last spectrum, whereas the adjacent continuum 
grows only by a factor 1.1. For the NaI~D lines, these two numbers are 1.4
and 1.2, still indicating a dichotomy between the pseudo-continuum and the
emission lines. But the most striking behaviour is that displayed by the
[OI]6300,6364\AA\/ doublet: the integrated flux changes by a factor
5.2 in about two hours and can be easily identified
as the responsible for the brightening observed in the $R$ passband.
This is easily visible in Figure~\ref{fig:skyevol}, where the [OI]6300\AA\/
component surpasses the [OI]5577\AA\/ in the transition from the first to 
the second spectrum and keeps growing in intensity in the subsequent two
spectra. The existence of these abrupt changes is known since the
pioneering work by Barbier (\cite{barbier}), who has shown that 
[OI]6300,6364\AA\/
can undergo strong brightness enhancements over an hour or two on two 
active regions about 20$^\circ$ on either side of the geomagnetic equator, 
which roughly corresponds to tropical sites. With Cerro Paranal included in 
one of these active areas, such events are not unexpected. A possible physical
 explanation for this effect is described by Ingham (\cite{ingham}), and 
involves the release of charged particles at the conjugate point of the 
ionosphere, which stream along the lines of force of the terrestrial magnetic
field. We notice that in our example, the first spectrum was taken about two 
hours before local midnight, at about one month before the end of austral 
summer. This is in contrast with Ingham's explanation, which implies that 
this phenomenon should take place in local winter, since in local summer the 
conjugate point, which for Paranal lies in the northern hemisphere, sees the 
sun later and not before, as it is the case during local winter.

Irrespective of the underlying physical mechanism, the [OI]6300,6364\AA\/ 
line intensity\footnote{Line intensities are here expressed in Rayleigh (R).
See Appendix \ref{sec:units}.}
changed from 255 R to 1330 R; the fact that the initial value is 
definitely higher than that expected at these geomagnetic latitudes ($<$50 R, 
Roach \& Gordon \cite{roachgordon}, Figure~4-12) seems to indicate that the line brightening 
had started before our first observation. On the other hand, the intensity of
the [OI]5577\AA\/ line in the first spectrum is 220 R, i.e. well in agreement
with the typical value (250 R, Schubert \& Walterscheid 2000). 

The case of NaI~D lines is slightly different, since these features follow a
strong seasonal variation which makes them brighter in winter and fainter in 
summer, the intensity range being 30-200 R (Schubert \& Walterscheid 2000).
This fluctuation is expected to produce a seasonal variation with an amplitude
of about 0.1 mag in the $R$ passband, while in $V$ the effect is negligible.
Actually, the minimum intensity of this feature
can change from site to site, according to the amount of light pollution. 
In fact, most of the radiation produced by low-pressure sodium lamps is 
released through this transition. For example, Benn \& Ellison (\cite{benn})
report for La Palma an estimated artificial contribution to the Sodium D
lines of about 70 R. In our first spectrum, the measured intensity is 
73 R, a value which, together with the epoch when it was obtained (end of 
summer) and the relatively large airmass ($X$=1.5), indicates a very small 
contribution from artificial illumination. However, a firmer limit can be 
set analysing a large sample of low resolution spectra taken around midsummer,
a task which is beyond the purpose of this paper.

\begin{figure}
\resizebox{\hsize}{!}{\includegraphics{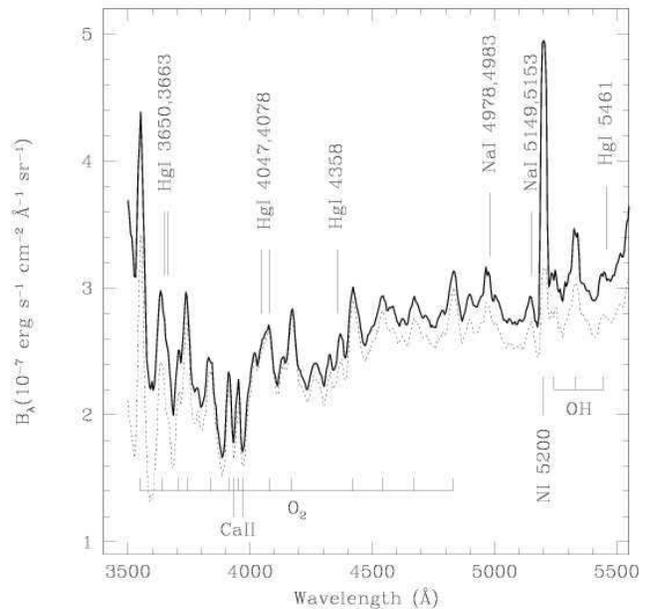}}
\caption{\label{fig:pollution} Night sky spectrum obtained at Paranal
on February 25, 2002 at 04:53 UT (see Figure~\ref{fig:skyevol}). Marked
are the expected positions for the most common lines produced by artificial
scattered light (upper ticks) and natural atmospheric features (lower ticks). 
The dotted line traces part of the spectrum taken during the same night at 
02:39 UT.
}
\end{figure}

To search for other possible signs of light pollution, we have examined
the wavelength range 3500-5500\AA\/ of the last spectrum presented in 
Figure~\ref{fig:skyevol}, which was obtained at a zenith distance of about
60$^\circ$ and at an azimuth of 313$^\circ$. 
A number of Hg and Na lines produced by street lamps, which are clearly 
detected at polluted sites, falls in this spectral region.
As expected, there is no clear trace of such features in the examined
spectrum; in particular, the strongest among these lines, HgI 4358\AA, is
definitely absent. This appears clearly in Figure~\ref{fig:pollution}, where
we have plotted the relevant spectral region and the expected positions 
for the brightest Hg and Na lines (Osterbrock \& Martel 1992). In the same
figure we have also marked the positions of O$_2$ and OH main features.
A comparison with the spectra presented by Broadfoot \& Kendall 
(\cite{broadfoot}) again confirms the absence of the HgI lines and shows that
almost all features can be confidently identified with natural transitions of
molecular oxygen and hydroxyl. There are probably two exceptions only, which
happen to be observed very close to the expected positions for NaI 4978,
4983\AA\/ and NaI 5149, 5163\AA, lines typically produced by high
pressure sodium lamps (Benn \& Ellison 1998). They are very weak,
with an intensity smaller than 2 R, and their contribution to the broad
band sky brightness is negligible. Nevertheless, if real, they
could indicate the possible presence of some artificial component in the
NaI~D lines, which are typically much brighter. This can be verified with
the analysis of a high resolution spectrum. If the contamination is really
present, this should show up with the broad components which are a clear
signature of high pressure sodium lamps. The inspection of a low airmass, 
high resolution (R=43000) and high signal-to-noise UVES spectrum of 
Paranal's night sky (Hanuschik et al. 2003, in preparation) has shown no 
traces of neither such broad components nor of other NaI and HgI lines.
For this purpose, suitable UVES observations at critical directions 
(Antofagasta, Yumbes mining plant) and high airmass periodically executed 
during technical nights, would probably allow one to detect much weaker 
traces of light pollution than any broad band photometric survey. But,
in conclusion, there is no indication for any azimuth dependency in our 
dark time $UBVRI$ measurements.

There are finally two interesting features shown in Figure~\ref{fig:pollution}
which deserve a short discussion. The first is the presence of CaII H\&K
absorption lines, which are clearly visible also in the spectra presented
by Broadfoot \& Kendall (\cite{broadfoot}) and are the probable result
of sunlight scattered by interplanetary dust (Ingham 1962). This is not
surprising, since the spectrum of Figure~\ref{fig:pollution} was taken at
$\beta$=$-$3\deg5 and $\lambda-\lambda_\odot$=139\deg8, i.e. in
a region were the contribution from the zodiacal light is significant
(see Figure~\ref{fig:zl}).
The other interesting aspect concerns the emission at about 5200\AA.
This unresolved feature, identified as NI, is extremely weak in the spectra of 
Broadfoot \& Kendall (\cite{broadfoot}), in agreement with its typical
intensity (1 R, Roach \& Gordon \cite{roachgordon}). On the contrary, in our first 
spectrum (dotted line in Figure~\ref{fig:pollution}) it is very clearly detected 
at an intensity of 7.5 R and steadily grows until it reaches 32 R in the 
last spectrum, becoming the brightest feature in this wavelength range.
This line, which is actually a blend of several very close NI transitions, is 
commonly seen in the Aurora spectrum with intensities of
0.1-2 kR (Schubert \& Walterscheid 2000) and it is supposed to originate in 
a layer at 258 km. The fact that its observed growth (by a factor 4.3) follows 
closely the one we have discussed for [OI]6300,6364\AA, suggests that the 
two regions probably undergo the same micro-auroral processes.

\begin{figure}
\resizebox{\hsize}{!}{\includegraphics{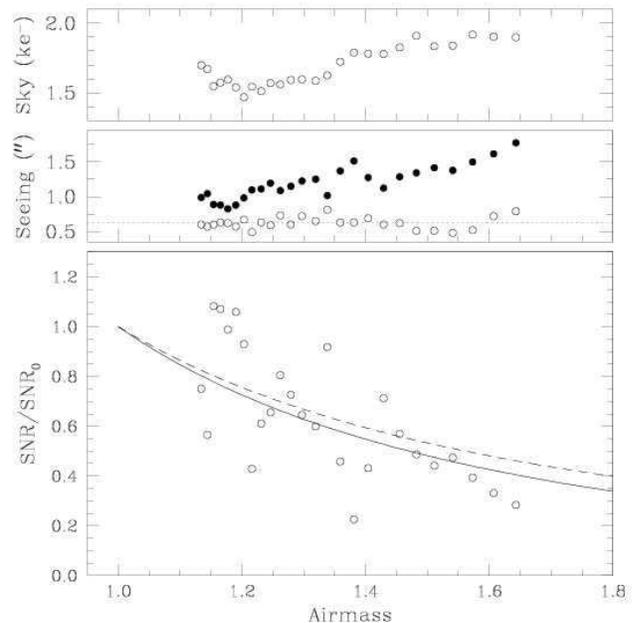}}
\caption{\label{fig:snr} Lower panel: peak signal-to-noise ratio measured
for the same star on a sequence of 150 seconds $I$ images obtained with
FORS1 on July 16, 2001. Solid and dashed lines trace Eq.~\ref{eq:snr} for
$U$ and $I$ passbands respectively. Middle panel: seeing measured 
by the Differential Image Motion Monitor (DIMM, Sandrock et al. 2000) at
5500 \AA\/ and reported to zenith (empty circles); each point represents 
the average of DIMM data over the exposure time of each image. The solid 
circles indicate the image quality (FWHM) directly measured on the images. 
Upper panel: sky background (in ke$^{-}$) measured on each image.
}
\end{figure}

Such abrupt phenomena, which make the sky brightness variations during a given
night rather unpredictable, are accompanied by more steady and well
behaved variations, the most clear of them being the inherent brightening
one faces going from small to large zenith distances.
In fact, as we have seen, the sky brightness increases at higher airmasses, 
especially in the red passbands, where it can change by 0.4 mag going from 
zenith to airmass $X$=2. For a given object, as a result of the photon shot 
noise increase, this turns into a degradation of the signal-to-noise ratio by 
a factor 1.6, which could bring it below the detection limit. Unfortunately, 
there are two other effects which work in the same direction, i.e. the 
increase of atmospheric extinction and seeing degradation. While the former 
causes a decrement of the signal, the latter tends to dilute a stellar image 
on a larger number of pixels on the detector.
Combining Eq.~\ref{eq:airmdep}, the usual atmospheric extinction law
$I=I_0\;10^{-0.4\kappa(X-1)}$ and the law which describes the variation of
seeing with airmass ($s=s_0\;X^{0.6}$, Roddier 1981) we can try to estimate
the overall effect on the expected signal-to-noise ratio at the central peak 
of a stellar object. After very simple calculations, one obtains the following
expression:

\begin{equation}
\label{eq:snr}
\frac{SNR(X)}{SNR_0}= X^{-1.2} \;
\left [ (1-f)+fX\right ]^{-\frac{1}{2}} \;
10^{-0.2\kappa(X-1)}
\end{equation}

where the 0 subscript denotes the zenith ($X$=1) value. Eq.~\ref{eq:snr} is
plotted in Figure~\ref{fig:snr} for the two extreme cases, i.e. $U$ and $I$
passbands. For comparison, we have overplotted real measurements performed
on a sequence of $I$ images obtained with FORS1 on July 16, 2001. 
Given the fact that the seeing was not constant during the sequence
(see the central panel), Eq.~\ref{eq:snr} gives a fair description of the observed
data, which show however a pretty large scatter. As one can see, the
average SNR ratio decreased by about a factor of 2 passing from airmass 
1.1 to airmass 1.6. Such degradation is not negligible, specially when
one is working with targets close to the detection limit. For this
reason we think that Eq.~\ref{eq:snr} could be implemented in the exposure
time calculators together with the model by Krisciunas \& Schaefer 
(\cite{krischae}), to allow for a more accurate prediction of the effective
outcome from an instrument. This can be particularly useful during
service mode observations, when now-casting of sky conditions at target's
position is often required.

\begin{acknowledgements}
We are grateful to K. Krisciunas and B. Schaefer for the discussion about 
the implementation of their model and to Bruno Leibundgut, Dave Silva,
Gero Rupprecht and Jean Gabriel Cuby for carefully reading the original
manuscript. We wish to thank Reinhard Hanuschik for providing us with the 
high resolution UVES night sky spectrum before publication. We are finally 
deeply indebted to Martino Romaniello, for the illuminating discussions, 
useful advices and stimulating suggestions.

All FORS1 images shown in this paper were obtained during Service
Mode runs and their proprietary period has expired.
\end{acknowledgements}

\appendix

\section{\label{sec:photcoeff} Extinction Coefficients and Colour Terms}

Extinction coefficients and colour terms were computed using the
method outlined by Harris, Pim Fitzgerald \& Cameron Reed (\cite{harris}),
i.e. via a single-step multi-linear least squares fit to the data
provided by the observation of photometric standard star fields
(Landolt 1992). This
procedure, in fact, allows one to get the photometric solutions using 
different stars with suitable colour and airmass ranges, without the need
for repeating the observations at different airmasses. Due to the FORS1
calibration plan implemented during the time range discussed in this work,
this was a mandatory requirement. To get meaningful results,
the method needs a fair number of stars well spread in colour and airmass.
While the first requirement is almost always fulfilled for the Landolt fields
suitable for CCD photometry, the second is more cumbersome to achieve using
FORS1 data.
In order to overcome this problem we have computed the photometric parameters
on a bi-monthly basis, for those time ranges where at least 3 stars were
observed at airmass larger than 1.6. Harris et al. (\cite{harris}) recommend
to use data spanning at least 1 airmass, but we had to relax this constraint
in order to get a sufficient number of stars.

\begin{table*}
\centering
\caption{\label{tab:ext} Bi-monthly average extinction coefficients 
(mag airmass$^{-1}$) measured at ESO-Paranal with FORS1 during
photometric nights. For each
entry the total number of data points ($n$) and the number of
data points with airmass larger than 1.6 ($n_z$) are reported.
The last row of the table shows the weighted means and the 
corresponding RMS errors.}
\tabcolsep 1.5mm
\begin{tabular}{cccccccccccccccc}
\hline \hline
Time &  \multicolumn{3}{c}{U} & \multicolumn{3}{c}{B} & \multicolumn{3}{c}{V} &
        \multicolumn{3}{c}{R} & \multicolumn{3}{c}{I} \\
Range & $\kappa$ &$n$&$n_z$ & $\kappa$ &$n$&$n_z$ & $\kappa$ &$n$&$n_z$ &$\kappa$ &$n$&$n_z$ &$\kappa$ &$n$& $n_z$\\
\hline
03/04-00 & 0.41$\pm$0.05& 87& 3&-             &  -& -&0.11$\pm$0.02 &120& 4&-             &  -& -&0.07$\pm$0.02&88 & 8\\
05/06-00 & -            &  -& -&0.25$\pm$0.01 & 43& 3&0.09$\pm$0.01 & 47& 4&0.07$\pm$0.01 & 46& 4&-            & - & -\\
07/08-00 & 0.47$\pm$0.01&122&12&0.23$\pm$0.01 &132&14&0.11$\pm$0.01 &162&19&0.07$\pm$0.01 &151&15&0.05$\pm$0.01&138&19\\  
11/12-00 & 0.43$\pm$0.01& 78& 9&0.21$\pm$0.01 & 86&16&0.11$\pm$0.01 & 89&13&0.07$\pm$0.01 & 83&14&0.03$\pm$0.01& 85&10\\
01/02-01 & 0.45$\pm$0.02&122&24&0.22$\pm$0.02 &154&29&0.10$\pm$0.01 &163&33&0.07$\pm$0.01 &148&28&0.04$\pm$0.01&145&29\\
05/06-01 & 0.40$\pm$0.03& 64& 4&0.19$\pm$0.02 & 66& 4&0.09$\pm$0.01 & 60& 3&0.06$\pm$0.01 & 60& 3&0.01$\pm$0.01& 68& 3\\  
07/08-01 & 0.43$\pm$0.02&117&18&0.22$\pm$0.01 &139&17&0.12$\pm$0.01 &143&20&0.08$\pm$0.01 &124&17&0.04$\pm$0.01& 68&17\\  
09/10-01 & 0.44$\pm$0.02&125& 3&0.24$\pm$0.01 &143& 3&0.11$\pm$0.01 &141& 4&0.08$\pm$0.01 &120& 4&0.06$\pm$0.01&130& 5\\
\hline
         & 0.43$\pm$0.02&   &&0.22$\pm$0.02 &   &&0.11$\pm$0.01 &   &&0.07$\pm$0.01 &   &&0.05$\pm$0.02& &\\
\hline
\end{tabular}
\end{table*}

Some tests have shown that the available photometric data do not allow the
computation of the second order colour term and extinction coefficient,
which are usually included in the photoelectric photometry solutions
(see Harris et al. 1991, their Eqs.~2.9).
In fact, the introduction of such terms in the fitting of our data produces
random oscillations in the solutions without any significant decrease in the
variance. Both second order coefficients are accompanied by large
errors and are always consistent with zero. This clearly means that the 
accuracy of our measurements is not sufficient to go beyond the first order
term. For this reason, we have adopted $M-m=m_0+\gamma\times 
C \; -\kappa\times z$ as fitting function
for the generic passband, where $M$ and $m$ are respectively the catalogue and 
instrumental magnitudes of the standard stars, $m_0$ is the zeropoint, $\gamma$ 
is the colour term with respect to some colour $C$, $\kappa$ is the extinction
coefficient and $z$ is the airmass. To eliminate clearly deviating measurements,
we have computed the photometric solutions in two steps. We have first used all
data to get a starting guess from which we have computed the global RMS deviation 
$\sigma$. Then we have rejected all data deviating more than 
1.5$\times \sigma$ and performed again the least squares fit on the selected
data. 

The results for the extinction coefficients are presented in Table~\ref{tab:ext}.
For each time range and passband we have reported the best fit value
of $\kappa$, the estimated RMS error (both in mag airmass$^{-1}$) and the 
number of data points used for the least squares fit. 

The values of $\kappa$ in the various bands show some minor fluctuations. Of 
course, we cannot exclude night-to-night variations, which are clearly 
observed at 
other sites (see for instance Krisciunas 1990). This would require a dedicated
monitoring, which is beyond the scope of this work and not feasible with the 
available data.
Here we can only conclude that, on average, the extinction coefficients
do not show any significant evolution or clear seasonal effects during the
20 months covered by the data discussed in this paper. Therefore, given also
the purpose of this analysis, we have assumed that the extinction coefficients
are constant in time and equal to the average values reported on
the last row of Table~\ref{tab:ext}.

The computed colour terms are shown Table~\ref{tab:cterm}. We note that in 
those cases where the constraints on the airmass range and the number of 
data-points were not fulfilled, we have performed the best fit keeping 
$\kappa$ constant and equal to the average values given in Table~\ref{tab:ext}. 
This prevented the best fit from giving results with no physical meaning due 
to the poor airmass coverage. As in the case of the extinction coefficients, 
the color terms show fluctuations which are within the errors, stronger
oscillations being seen in the $U$ passband, where the accuracy of the
photometry is lower.

On the basis of these results we can conclude that there are no strong 
indications for significant colour term evolution and it is therefore
reasonable to assume that they are constant in time.
For our purposes we have adopted the weighted mean values shown in the last
row of Table~\ref{tab:cterm}.

\begin{table*}
\centering
\caption{\label{tab:cterm} Bi-monthly average colour terms for FORS1. The 
total number of used data points are the same reported in Table~\ref{tab:ext},
while the estimated RMS uncertainty on each entry is $0.01$. 
The last row shows the weighted mean values and their estimated RMS errors.}
\tabcolsep 2mm
\begin{tabular}{cccccc}
\hline \hline
 Time Range & $\gamma_{U-B}^U$& $\gamma_{B-V}^B$& $\gamma_{B-V}^V$& $\gamma_{V-R}^R$& $\gamma_{V-I}^I$ \\
\hline
03/04-00 & 0.09&$-$0.07 &0.04 &0.02 &$-$0.05 \\
05/06-00 & 0.08&$-$0.08 &0.04 &0.02 &$-$0.04 \\
07/08-00 & 0.10&$-$0.08 &0.03 &0.00 &$-$0.05 \\
09/10-00 & 0.08&$-$0.07 &0.05 &0.04 &$-$0.04 \\
11/12-00 & 0.04&$-$0.09 &0.04 &0.04 &$-$0.04 \\
01/02-01 & 0.07&$-$0.08 &0.03 &0.03 &$-$0.05 \\
03/04-01 & 0.07&$-$0.08 &0.04 &0.03 &$-$0.04 \\
05/06-01 & 0.03&$-$0.09 &0.05 &0.02 &$-$0.04 \\
07/08-01 & 0.10&$-$0.08 &0.03 &0.02 &$-$0.05 \\
09/10-01 & 0.08&$-$0.08 &0.04 &0.04 &$-$0.03 \\
\hline
         & 0.07$\pm$0.02&$-$0.08$\pm$0.01 &0.04$\pm$0.01 &0.03$\pm$0.01 &$-$0.04$\pm$0.01 \\
\hline
\end{tabular}
\end{table*}

\section{\label{sec:units} Sky brightness units}

Throughout this paper we have expressed the sky brightness in
mag arcsec$^{-2}$. Since other authors have used different units, we derive
here the conversions for the most used ones.
In the $cgs$ system, the sky brightness is expressed in erg s$^{-1}$ cm$^{-2}$
\AA$^{-1}$ sr$^{-1}$. Now, if $m_{sky,\lambda}$ is the sky brightness in a 
given passband (in mag arcsec$^{-2}$) and $m_{0,\lambda}$ is the photometric
system zeropoint of that passband, the conversion is obtained as follows:

\begin{equation}
\label{eq:convgcs}
B_\lambda (cgs)= 10^{-0.4(m_{sky,\lambda}-m_{0,\lambda}-26.573)}
\end{equation} 

where the constant in the exponent accounts for the fact that 
1 arcsec$^2$ corresponds to 2.35$\times$10$^{-11}$~sr. For example, if we use
the constant $m_0$ for the Johnson system $V$ band given by Drilling \& 
Landolt (\cite{drilling}), for a typical night sky brightness 
$m_{sky,V}$=21.6 mag arcsec$^{-2}$, this gives 
$B_V\simeq$3.7$\times$10$^{-7}$ erg s$^{-1}$ cm$^{-2}$ \AA$^{-1}$ sr$^{-1}$. 
For practical reasons, we have introduced the surface brightness unit sbu (see 
Sec.~\ref{sec:survey}), which is defined 
as 1 sbu$\equiv$10$^{-9}$ erg s$^{-1}$ cm$^{-2}$ \AA$^{-1}$ sr$^{-1}$. With this 
setting, the typical $V$ sky brightness is 366 sbu.

Often $B_\lambda$ is expressed in $SI$ units. The conversion from $cgs$ units
is simple, since 1 erg s$^{-1}$ cm$^{-2}$ \AA$^{-1}$ sr$^{-1}$$\equiv$
10 W m$^{-2}$ sr$^{-1}$ $\mu$m$^{-1}$ and hence:

\begin{equation}
\label{eq:convsi}
B_\lambda (SI)= 10^{-0.4(m_{sky,\lambda}-m_{0,\lambda}-29.073)}
\end{equation} 

so that the typical $V$ sky brightness turns out to be 3.7$\times$10$^{-6}$
W m$^{-2}$ sr$^{-1}$ $\mu$m$^{-1}$.

Especially in the past, the sky brightness was expressed in
$S_{10}$, which is defined as the number of 10$^{th}$ magnitude stars
per square degree required to produce a global brightness equal to
the one observed in the given passband. Since 1 sr corresponds to 
3.282$\times$10$^3$ square degrees, the conversion is given by the 
following equation:

\begin{equation}
\label{eq:s10}
1 \; S_{10}(\lambda) = 10^{-0.4(1.21-m_{0,\lambda})}\;\;
\mbox{erg s$^{-1}$ cm$^{-2}$ \AA$^{-1}$ sr$^{-1}$}
\end{equation}

Again, using the constant given by Drilling \& Landolt (\cite{drilling}) 
for the $V$ band, one obtains 
$1\;S_{10}(V)\equiv$1.24$\times$10$^{-9}$ erg s$^{-1}$ cm$^{-2}$ 
\AA$^{-1}$ sr$^{-1}$ $\equiv$ 1.24 sbu. For $U$, $B$, $R$ and $I$ the 
multiplicative constant is 1.38, 2.11, 0.57 and 0.28, respectively.
Therefore, the sky brightness expressed in $S_{10}$ and in sbu are of the 
same order of magnitude in all optical broad bands. For example, the typical 
$V$ sky brightness is 295 $S_{10}(V)$. 

Another unit used in the past is the nanoLambert (nL), which measures the 
{\it perceived} surface brightness and corresponds to 3.80 $S_{10}(V)$. For 
the $V$ passband this gives also $1\;nL \equiv 4.72$ sbu.

Finally, to quantify the intensity of night sky emission lines, the Rayleigh (R)
unit is commonly adopted. It is defined as 10$^6$/4$\pi$ photons s$^{-1}$ 
cm$^{-2}$ sr$^{-1}$ and the conversion from erg s$^{-1}$ cm $^{-2}$ sr$^{-1}$
is given by the following expression:

\begin{equation}
\label{eq:rayleigh}
1R \equiv 634.4 \; \lambda \; \mbox{erg s$^{-1}$ cm$^{-2}$ sr$^{-1}$}
\end{equation}

where $\lambda$ is expressed in \AA. The intensity of non-monochromatic 
features, like the pseudo-continuum, is generally expressed in R \AA$^{-1}$.

For a more thorough discussion of the sky brightness units the reader is 
referred to Leinert et al. (\cite{leinert98}) and Benn \& Ellison 
(\cite{benn}).

\section{\label{sec:airmdep} Airmass dependency}

The results obtained by Garstang (\cite{garstang89}) can be used to derive
an approximate expression for the sky brightness dependency on the zenith
distance, as already pointed out by Krisciunas \& Schaefer (\cite{krischae}).
If we assume that a fraction $f$ of the total sky brightness is 
generated by the airglow and the remaining $(1-f)$ fraction is produced
outside the atmosphere (hence including zodiacal light, faint stars and
galaxies), Garstang's Eq.~29 can be rewritten in a more
general way as follows:

\begin{equation}
\label{eq:garst}
b(Z)=b_0\;[ (1-f) + f\;X)]
\end{equation}

where $X$, the optical pathlength along a line of sight, is given by

\begin{equation}
\label{eq:X}
X=(1-0.96\;sin^2Z)^{-1/2}
\end{equation}

$Z$ being the zenith distance. For the hypothesis on which this result
is based, the reader is referred to Garstang's original pubblication and
to the analysis presented by Roach \& Meinel (\cite{roach}). Here we just
recall that Eqs.~\ref{eq:garst} and \ref{eq:X} were obtained assuming that
the {\it extra-terrestrial} fraction $(1-f)$ comes from infinity, while
the airglow emission is generated in a layer (the so called {\it 
van Rhjin layer}) placed at 130 km from the Earth's surface.

While the intensity of the extra-terrestrial component is independent on
the zenith distance, this is not true for the airglow, due to the variable
depth of the emitting layer along the line of sight. This is taken into
account by the term $X$, which grows towards the horizon.

In his model Garstang (\cite{garstang89}) considers three different mechanisms
for the background light to reach the observer once it leaves the van
Rhjin layer: direct transmission, aerosol scattering and Rayleigh
scattering. As a matter of fact, the first channel dominates on the others
(see Figure~5 in Garstang 1989) which, in a first approximation, can be 
safely neglected. One is therefore left with the propagation of the sky
background light across the lower atmosphere, which is described by 
Garstang's Eq.~30 through the extinction factor $EF$. Now, if $\kappa$
is the extinction coefficient for a given passband (in mag airmass$^{-1}$), 
we can write $EF\simeq 10^{-0.4\;(X-1)}$, so that the expected change
in the sky brightness as a function of $X$ is given by:

\begin{equation}
\label{eq:airmdep}
\Delta m = -2.5\;log [(1-f)+f\;X] + \kappa \; (X-1)
\end{equation}

We notice that using $f$=1 in this equation gives Eq.~2 of Krisciunas \& 
Schaefer (\cite{krischae}), which gives slightly larger values for $\Delta m$
than Eq.~\ref{eq:airmdep}.

\begin{figure}
\resizebox{\hsize}{!}{\includegraphics{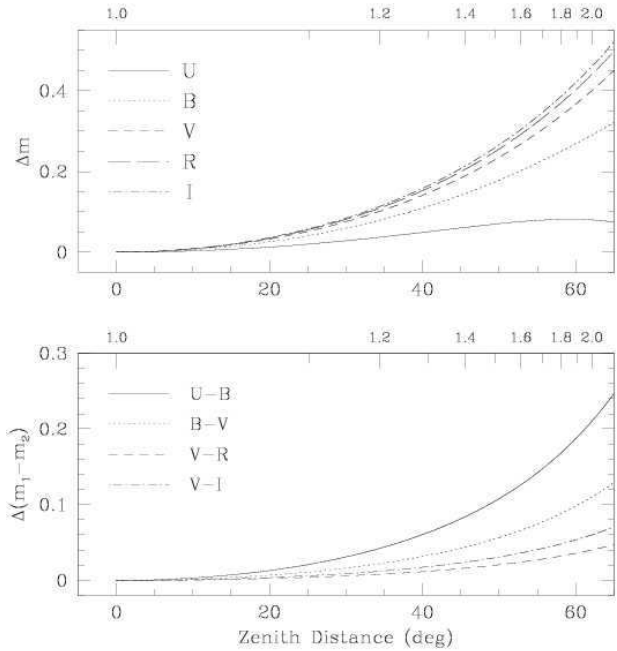}}
\caption{\label{fig:airmdep} Upper panel: sky brightness variation
(in mag arcsec$^{-2}$) as a function of zenith distance expected 
from Eq.~\ref{eq:airmdep} for $f$=0.6. For each passband the mean Paranal 
extinction coefficients presented in Table~\ref{tab:ext} were adopted. Lower
panel: expected colour variation as a function of zenith distance. In both 
plots the upper scale reports the optical pathlength $X$, expressed in
airmasses.
}
\end{figure}

The expected behaviour is plotted in Figure~\ref{fig:airmdep}, where we
present the predicted sky brightness (upper panel) and sky colour variations
(lower panel) for the different passbands. For $f$ we have assumed a typical
value of 0.6 (see Garstang 1989 and Benn \& Ellison 1998). As one can see 
the correction becomes relevant for$Z\geq$30$^\circ$ and reaches values 
as large as $\sim$0.4 mag at $Z$=60$^\circ$
(see also Table~\ref{tab:airmdep}). These predictions are consistent
with the measurements obtained by Pilachowski et al. (\cite{pila}), 
Mattila et al. (\cite{attila}), Leinert et al. (\cite{leinert95}) and 
Benn \& Ellison (\cite{benn}). In particular, Pilachowski et al. (\cite{pila})
have clearly detected the sky inherent reddening (see their Figure~3), which
is in fair agreement with the predictions of Eq.~\ref{eq:airmdep} for the
$(B-V)$ colour (see also Figure~\ref{fig:airmdep}, lower panel).

\begin{table}
\centering
\caption{\label{tab:airmdep} Zenith corrections (mag) computed according to
Eq.~\ref{eq:airmdep}. The values have been calculated using 
Paranal average extinction coefficients $\kappa$ (see Tab~\ref{tab:ext})
and $f$=0.6. The zenith distance $Z$ is expressed in degrees, while $X$ 
is in airmasses.}
\begin{tabular}{ccccccc}
\hline \hline
 $Z$ & $X$  &$\Delta U$&$\Delta B$&$\Delta V$&$\Delta R$&$\Delta I$ \\
\hline
10.0 & 1.01 & 0.00 & 0.01 & 0.01 & 0.01 & 0.01\\
20.0 & 1.06 & 0.01 & 0.03 & 0.03 & 0.03 & 0.04\\
30.0 & 1.15 & 0.03 & 0.06 & 0.08 & 0.08 & 0.08\\
40.0 & 1.29 & 0.05 & 0.11 & 0.14 & 0.15 & 0.16\\
50.0 & 1.51 & 0.07 & 0.18 & 0.24 & 0.26 & 0.27\\
60.0 & 1.89 & 0.08 & 0.27 & 0.37 & 0.40 & 0.42\\
\hline
\end{tabular}
\end{table}

\begin{figure}
\resizebox{\hsize}{!}{\includegraphics{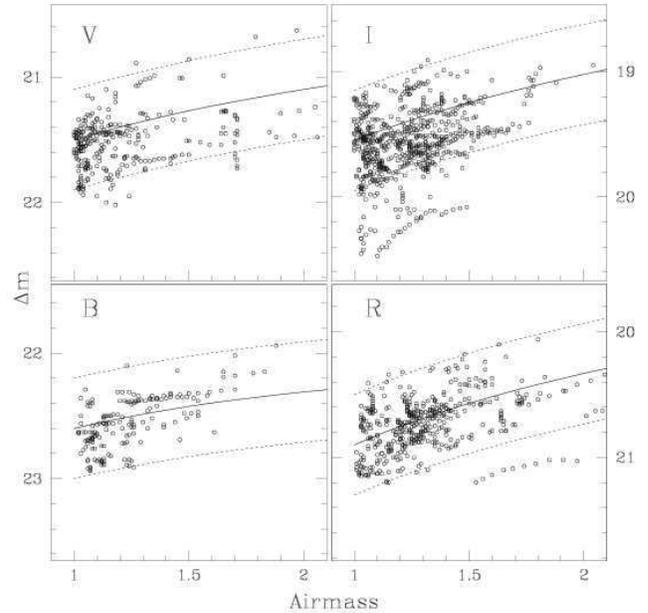}}
\caption{\label{fig:airmtrend} $B,V,R$ and $I$ sky brightness during
dark time as a function of airmass. The solid line traces 
Eq.~\ref{eq:airmdep}, while the dotted lines are placed at $\pm$0.4
mag from this law.
}
\end{figure}

\begin{figure}
\resizebox{\hsize}{!}{\includegraphics{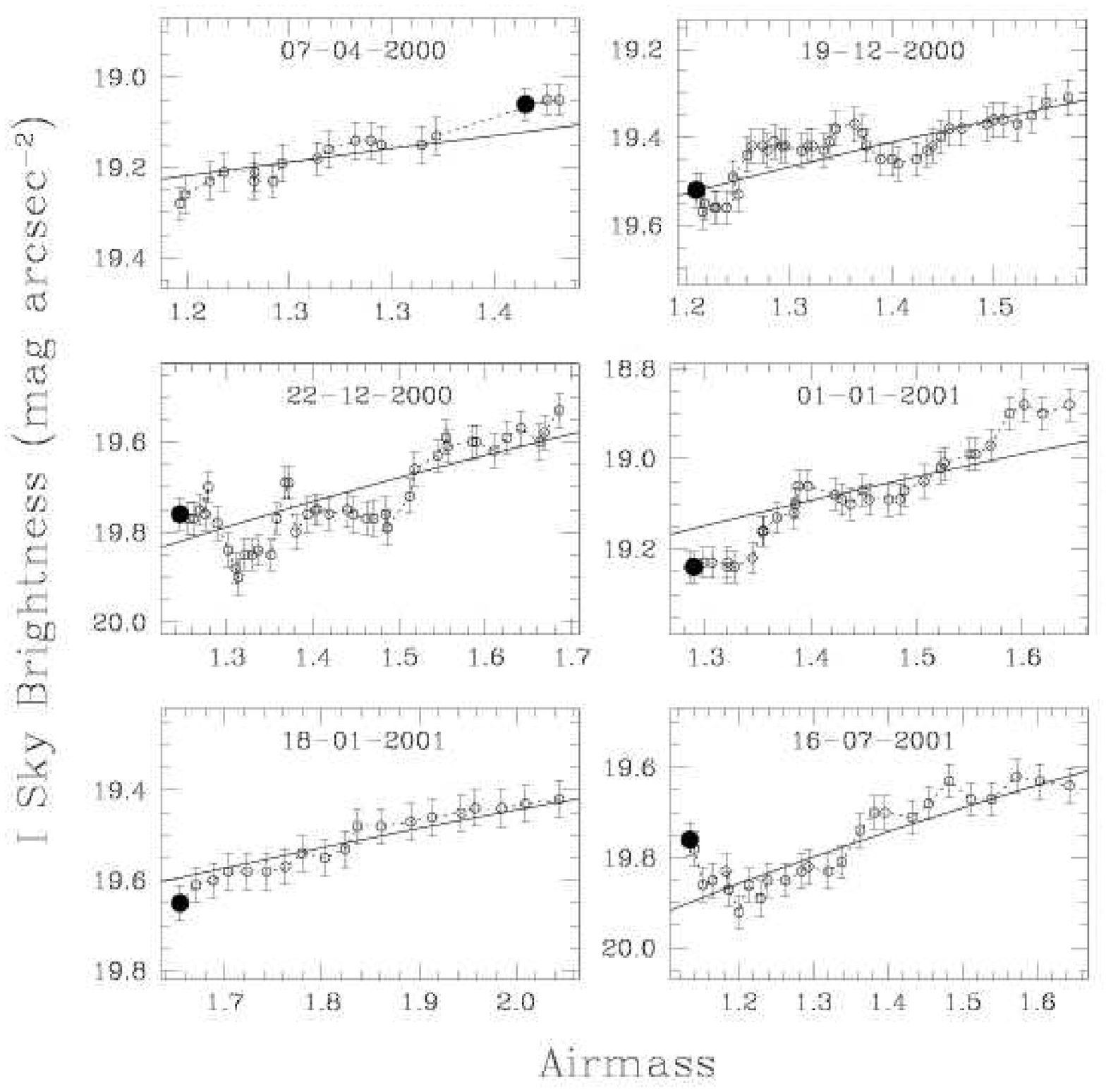}}
\caption{\label{fig:examples1} Sky brightness sequences on six different 
moonless nights at Paranal in the $I$ passband, corrected for zodiacal
light contribution. 
The dotted line connects chronologically the various data points, while
the filled dot indicates the first observation in each time series. Finally,
the solid line traces Eq.~\ref{eq:airmdep} with $f$=0.7. 
}
\end{figure}

Our data confirm these findings, as it is shown in Figure~\ref{fig:airmtrend},
where we have plotted Paranal dark time measurements for $B, V, R$ and $I$
passbands. Even though the scattering due to night to night variations is 
quite large, the overall trend is consistent with the predictions of 
Eq.~\ref{eq:airmdep}. The $U$ data, which reach airmass $X$=1.36 only, do 
not show a clear sky brightness increase, and this is again compatible with 
the expected value ($\sim$0.05 mag).

The airmass dependency is better visible when considering data obtained
during single nights, when the overall sky brightness is reasonably stable.
This is illustrated in Figure~\ref{fig:examples1}, where we present 
six very good series all obtained in the $I$ passband during moonless nights 
at Paranal. For comparison, we have plotted Eq.~\ref{eq:airmdep}
with $f$=0.7, which reproduces fairly well the observed trend from 
airmass 1.2 to 2.0. As one can see, smooth deviations with peak values 
of $\sim$0.1 mag arcsec$^{-1}$ are detected within each single
night, while the range spanned by the zenith extrapolated values is as 
large as $\sim$0.7 mag.
The fact that a comparably large dispersion is seen in the other filters
(see. Figure~\ref{fig:airmdep}) suggests that night-to-night variations
of this amplitude are common to all optical passbands.

We note that on April 7, 2000 and January 1, 2001 the observed slope was 
higher than predicted by Eq.~\ref{eq:airmdep}. This implies that the 
contribution by the airglow to the global sky brightness in those nights was 
probably larger than the value we have used here, i.e. 70\%. This hypothesis 
is in agreement with the fact that on 07-04-2000 and 01-01-2001 the 
extrapolated zenith value was 0.4$-$0.5 mag brighter than in all other 
nights shown in Figure~\ref{fig:examples1}.

\section{\label{sec:walker} The Walker effect revisited}

\begin{figure}
\resizebox{\hsize}{!}{\includegraphics{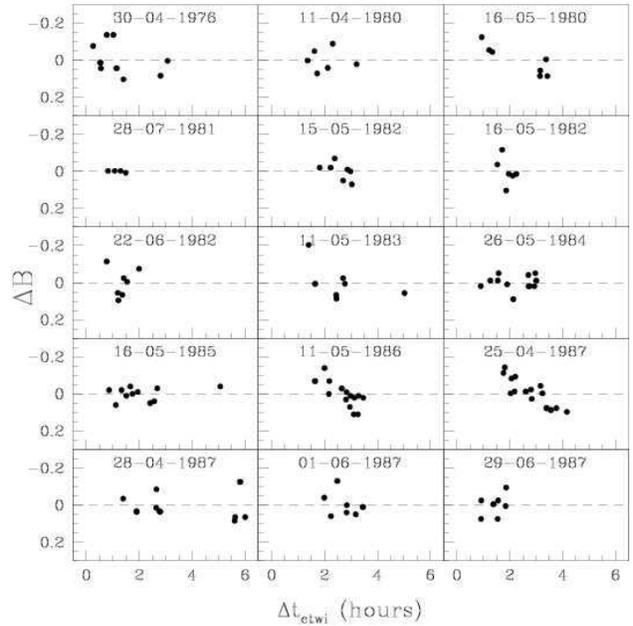}}
\caption{\label{fig:seqwalker} Zenith sky brightness at San Benito
Mountain in the $B$ passband from Walker (\cite{walker88b}). For clarity, 
the mean sky brightness has been subtracted to each time series.}
\end{figure}

In this section we discuss and re-examine the findings published by 
Walker (\cite{walker88b}).
In Figure~\ref{fig:seqwalker} we have plotted the $B$ data of his Table~1,
with the exception of July 8, 1980 and May 4, 1984 (since they cover a very
small time range with 2 and 3 points only) and the two data points marked 
by Walker with an asterisk in the sequences of May 11, 1983 and Apr 28, 1987.
As one can see, there are only two dates in which a clear decreasing
trend is visible, i.e. May 11, 1986 and April 25, 1987. Linear squares
fitting to the two data sets give slopes of $\sim$0.08 $\pm$0.02 
mag hour$^{-1}$ both in $B$ and $V$ filters, which turn into a sky
brightness decrease of 0.48$\pm$0.12 mag during the first 6 hours of 
the night. Since Walker's data have been collected across a full sunspot
cycle, possible systematic variations due to the solar activity
have to be removed. This can be achieved shifting all time
sequences in order to have the same sky brightness at some reference
time, an operation that also has the effect of correcting for night-to-night
variations in the overall sky brightness. 
For this purpose we have performed, for each date, a linear
least squares fit to the data and we have interpolated the resulting
straight line at a time distance from the evening twilight 
$\Delta t_{etwi}$=2 hours. 
We note that a similar procedure must have been followed by Walker 
(\cite{walker88b}), since in his Figure~2 all measurements refer to the
magnitude at the end of evening twilight ($\Delta t_{etwi}$=0). Even though
not explicitly mentioned in the paper, this implies that some 
extra/interpolation had to be performed, since estimates at 
$\Delta t_{etwi}$=0 are practically never available.

\begin{figure}
\resizebox{\hsize}{!}{\includegraphics{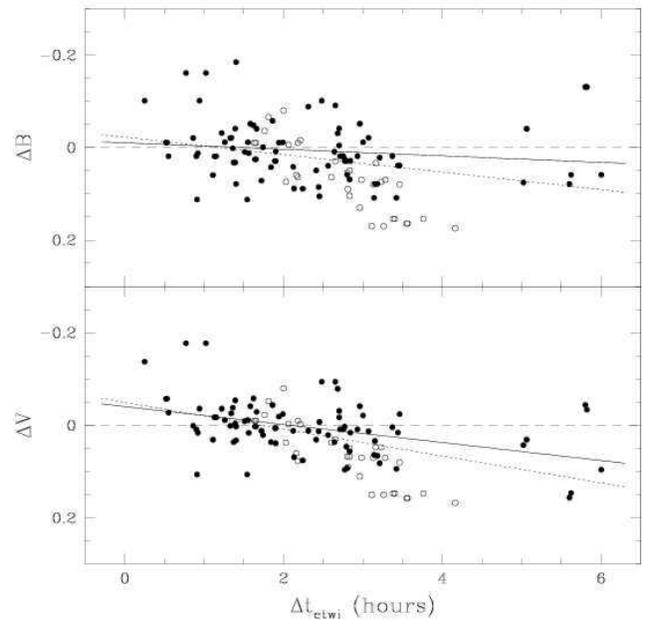}}
\caption{\label{fig:rwalker} Variation in zenith sky brightness at San
Benito Mountain with time after the end of astronomical twilight. Data
have been corrected for differential zodiacal light contribution and
all time series were normalised to the interpolated brightness at 
$\Delta t_{etwi}=2$ hours. Dotted and solid lines trace a linear least squares
fitting with and without the data of May 11, 1986 and April 25, 1987
respectively (empty circles). Data are from Walker (\cite{walker88b}).}
\end{figure}

To avoid meaningless extrapolations we have disregarded the data from 
July 28, 1981, which do not cover a suitable time range. Following Walker, 
we also have not included the data obtained in 1982, since they were most 
likely affected by the Chinchonal volcano eruption. 
Finally, to account for the differential contribution of zodiacal light,
we have corrected Walker's measurements using the data by
Levasseur-Regourd \& Dumont  (\cite{levasseur}) and assuming that all 
observations were carried out at the zenith of the observing site.
The results one obtains following this procedure are presented in 
Figure~\ref{fig:rwalker}, which shows a much less convincing evidence for
a systematic trend than Walker's Figure~2. In that case in fact, a 
decrement of $\sim$0.4 mag is seen during the first four hours, a 
behaviour which is definitely not visible in our Figure~\ref{fig:rwalker}. 
As a matter of fact, a linear least squares fitting gives a rate of 
0.02$\pm$0.01 and 0.03$\pm$0.01 mag hour$^{-1}$ for $B$ and $V$ respectively,
which reduce to 0.01$\pm$0.01 and 0.02 $\pm$0.01 mag hour$^{-1}$ if
one excludes the two sequences of May 11, 1986 and April 25, 1987
(empty circles). From these values we can deduce
a maximum decrement of 0.1 and 0.2 mag in $B$ and $V$ respectively
during the first 6 hours after the end of twilight. Therefore, our
conclusion is that even though there is an indication for a systematic
trend in Walker's data, the rate is significantly lower than was thought.
Moreover, due to the limited number of nights and the small time coverage
of several of the time series, the results one gets may depend on the
behaviour recorded during a few well sampled nights. For example,
all data at $\Delta t_{etwi}>$5.5 hours were collected on April 28, 1987.
This could explain why no other author has found the same strong effect
when using larger data bases (see for instance Figure~9 of 
Benn \& Ellison 1998).

\end{document}